\documentclass[journal]{IEEEtran}
\usepackage[utf8x]{inputenc}
\usepackage[english]{babel}
\usepackage{verbatim} 
\usepackage{varioref}
\usepackage{subfigure} 
 
\usepackage{psfrag}
\usepackage{epsf}
\usepackage{graphics}
\usepackage{graphicx}
\usepackage{amsbsy}
\usepackage{amssymb}
\usepackage{amscd}
\usepackage{amsmath}
\usepackage{amsthm}
\usepackage{enumerate} 
\usepackage{color}
\usepackage{epstopdf}

\usepackage{threeparttable}
\usepackage{algorithm2e}

\usepackage{cite}
\usepackage{url}

\newcommand{\overbar}[1]{\mkern 1.5mu\overline{\mkern-1.5mu#1\mkern-1.5mu}\mkern 1.5mu}

\def\bA{{\boldsymbol{A}}}
\def\bP{{\boldsymbol{P}}}

\def\bb{{\boldsymbol{b}}}

\def\bD{{\boldsymbol{D}}}

\def\bR{{\boldsymbol{R}}}

\def\bW{{\boldsymbol{W}}}

\title{Gbps User Rates Using mmWave Relayed Backhaul with High Gain Antennas}

\author{
Jinfeng Du, \emph{Member, IEEE}, 
Efe Onaran, \emph{Student Member, IEEE},
Dmitry Chizhik, \emph{Fellow, IEEE}, Sivarama Venkatesan,
and Reinaldo A. Valenzuela, \emph{Fellow, IEEE}
\thanks{J. Du, D. Chizhik, S. Venkatesan, and R. A. Valenzuela are with Nokia Bell Labs, Holmdel, NJ 07733, USA. Email: 
\{jinfeng.du, dmitry.chizhik, venkat.venkatesan, reinaldo.valenzuela\}@nokia-bell-labs.com. }
\thanks{E. Onaran was a summer intern at Nokia Bell Labs,  Holmdel, NJ 07733, USA. He is with NYU Tandon School of Engineering, Brooklyn, NY 11201, USA (Email: eonaran@nyu.edu).}
} 

\begin{document}

\maketitle

\begin{abstract}

Delivering Gbps high user rate over long distances ($\sim$1 km) is challenging, and the abundant spectrum available in millimeter wave band cannot solve the challenge by its own due to the severe path loss and other limitations. Since it is economically challenging to deploy wired backhaul every few hundred meters, relays (e.g., wireless access points) have been proposed to extend the coverage of a base station which has wired connection to the core network. These relays, deployed every few hundred meters, serve the users in their vicinity and are backhauled to the base station through wireless connections. In this work,   the wireless relayed backhaul design has been formulated as a topology-bandwidth-power joint optimization problem, and the influence of path loss, angular spread,  array size, and RF power limitation on the user rate has been evaluated. It has been shown that for a linear network deployed along the street at 28 GHz, when high joint directional gain (50 dBi) is available, 1 Gbps user rate within cell range of 1 km can be delivered using 1.5 GHz of bandwidth (using single polarization antennas). The user rates drop precipitously when joint directional gain is reduced, or when the path loss is much more severe. When the number of RF chains is limited, the benefit of larger arrays will eventually be surpassed by the increased channel estimation penalty as the effective beamforming gain saturates owing to the channel angular spread.  
\end{abstract}
 

\section{Introduction}\label{sec:introduction}

     Very wide spectrum (on the order of GHz, as permitted by FCC~\cite{fcc}) available in millimeter wave (mmWave) bands provides great relief on the shortage of signaling bandwidth in traditional cellular networks. However, the high path loss, sensitivity to blockage and foliage, RF hardware limitation, and other difficulties at high frequencies make it challenging to provide high user rate without shrinking the traditional cell coverage range. When channel estimation penalty\cite{Hassibi-Hochwald} is accounted for, using wider bandwidth may be ineffective or even counterproductive, as revealed in~\cite{BWopt},  for transmissions at range beyond a few hundred meters and in non-line-of-sight (NLOS) scenarios, where the system is usually noise limited. 
	
		On the other hand, it is economically challenging to deploy wired backhaul every few hundred meters, which limits the deployment of ultra dense networks that are critical to support high user rates in mmWave bands~\cite{Pi2011, Rangan2014, Ghosh2014}. As a viable alternative,  relays (e.g., wireless access points) have been proposed to extend the coverage of a base station which has wired connection to the core network. 
		Deploying relays in wireless communication systems is a well-studied concept~\cite{maric2004bandwidth, reznik2002capacity, islam2012optimal, bakanoglu2011resource, liang2004resource, liang2007resource}, where different bandwidth and power allocation techniques have been evaluated for a given relay network setup/topology, using  Decode-and-Forward (DF) relays~\cite{ liang2007resource, gong2010joint} or Amplify-and-Forward (AF)  relays~\cite{zhao2006improving, gao2008channel}  in wireless fading environment. 
In recently years relays using mmWave backhaul have attracted much attention~\cite{Hur2013, Taori2015,Singh2015,biswas2016performance} thanks to the high beamforming gain supported by large arrays. In~\cite{Hur2013} a beam alignment scheme based on subspace sampling and layered beamforming codebooks is proposed to balance the array gain and beam misalignment loss from wind-induced movement.
Performance and challenges of deploying in-band	relays in dense networks are studied in \cite{Taori2015}.	The use of  relays to combat blockage and to enhance  coverage is studied in~\cite{biswas2016performance} where the randomly deployed relays are modeled by independent Poisson point process (PPP).   In~\cite{Singh2015} only a fraction of base stations are equipped with wired backhaul and serve  the nearby non-wired base stations using mmWave links.  By modeling base station positions as PPP, coverage and rate distribution have been characterized.

   In this work we consider the use of relays in mmWave bands to support Gbps user rate at traditional cell coverage range, e.g., around 1 km.  Relays are deployed a few hundred meters apart and  these relays serve the users in their vicinity and are backhauled to the wired base station through wireless links. 	One of the appealing deployment scenarios is the street canyon, as illustrated in Figure~\ref{fig:topology}, where  relays are deployed along the street and are connected  to  the base station using  wireless backhaul. By adding 4 such  relays, 200 meters apart,  on one side of the base station, the traditional cell coverage range of 1 km can be approached. Such arrangement can be repeated along street grids in urban areas for coverage.  The direct backhaul links between the base station and relays  illustrated in Figure~\ref{fig:topology} (a) is just one example out of many possible ways to provide high speed wireless backhaul. Using high gain directional antenna arrays, it is possible to explore the relay-to-relay mmWave links for backhaul, as  illustrated in Figure~\ref{fig:topology} (b) and (c). How to optimally design the backhaul topology and allocate bandwidth/power across all wireless links is the problem we are going to address in this work so as to maximize the cell throughput.

To highlight the essence of the work, in the rest of this paper, we focus on the street canyon deployment scenario as illustrated in Figure~\ref{fig:topology} to present our design of the wireless backhaul in supporting Gbps user rates at cell edge. We formulate an optimization problem that determines jointly the optimal backhaul topology, bandwidth allocation, and power split among different links. The influence of path loss, antenna gains, angular spread, array size, and RF power limitation are evaluated to identify the feasible operation regimes the sustain high user rates at cell edges. Although the results presented here are customized for linear networks, which is most suitable for street canyon scenarios, our analysis framework and the optimization setup can be extended straightforwardly to more general relay networks.

The rest of this paper is organized as follows. We present the system model in Section~\ref{sec:model} and the joint optimization in Section~\ref{sec:optimization} to determine the optimal backhaul topology and corresponding bandwidth and power allocation.  In Section~\ref{sec:angular} we investigate the effect of channel angular spread on the effective beamforming gain. Numerical results are presented in Section~\ref{sec:num} to evaluate the  influence of path loss, antenna gains, angular spread, array size, and RF power limitation. Conclusions are in Section~\ref{sec:conclusion}.

\section{System Models}\label{sec:model}

\begin{figure}[t]
		\centering
		\includegraphics[width=0.96\columnwidth]{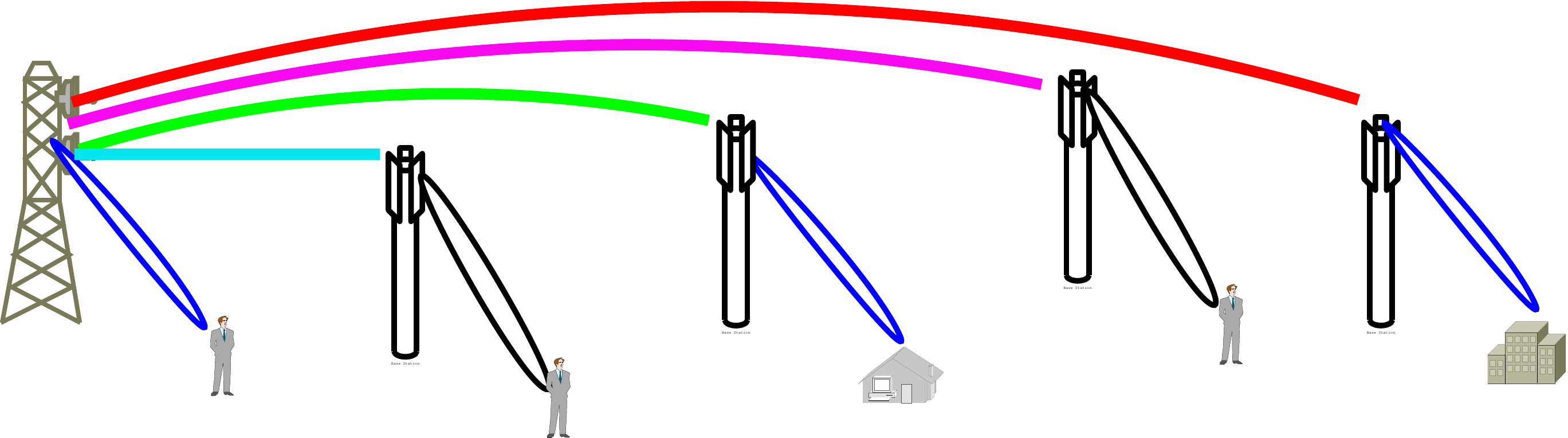} \hspace{ 3mm} \emph{(a) Single-Hop}\\
		\includegraphics[width=0.96\columnwidth]{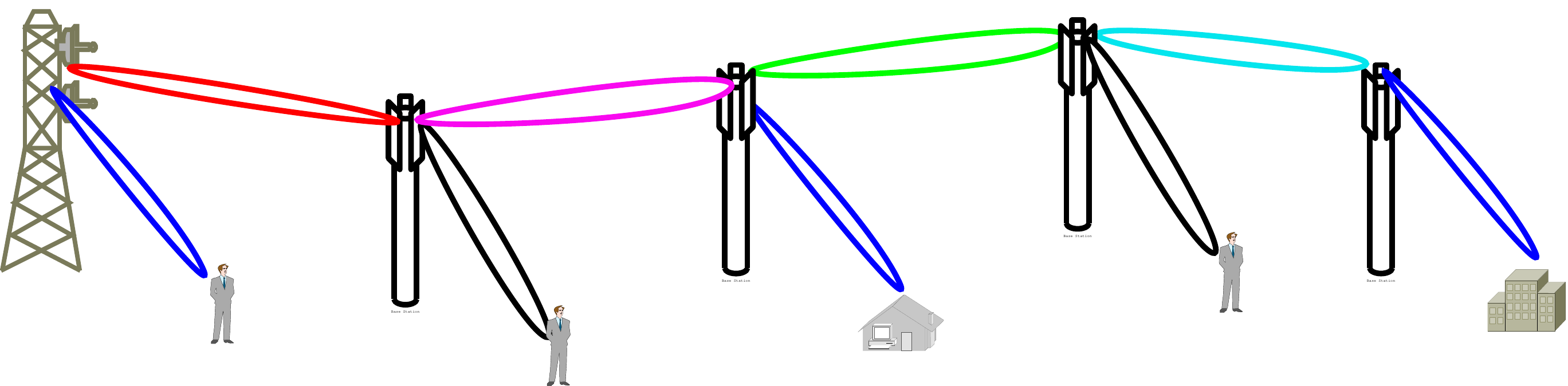}\hspace{ 5mm} \emph{(b) Nearest Neighbor}\\
		\includegraphics[width=0.96\columnwidth]{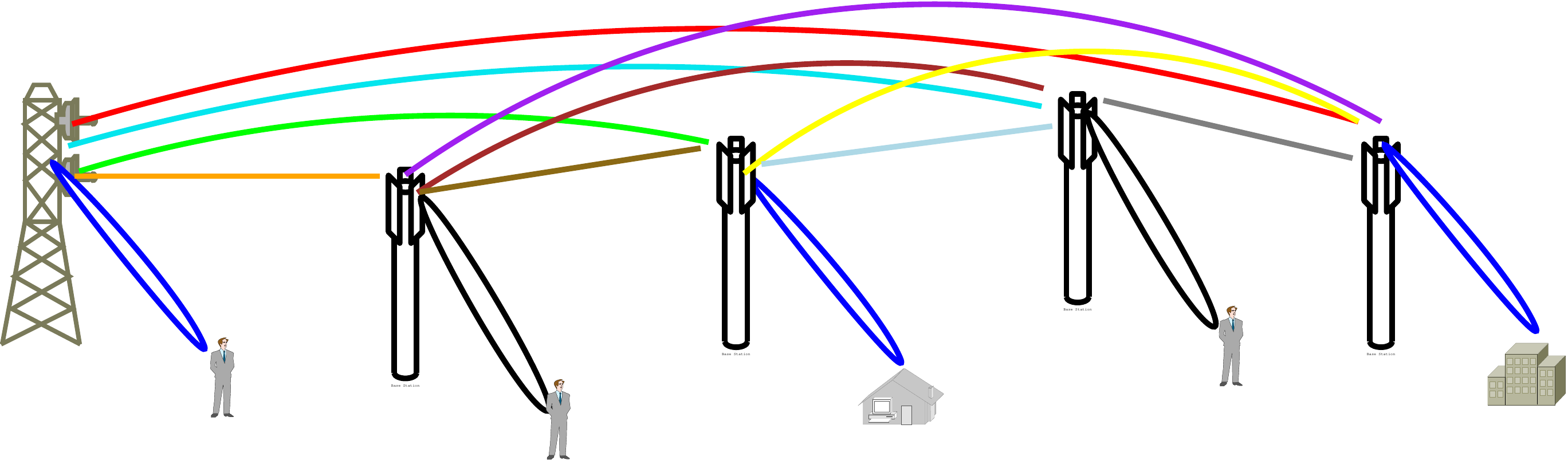} \hspace{ 3mm} \emph{(c) full-connectivity}
		\caption{Illustration of wireless backhaul where relays are deployed along a street and backhauled to the wired access point (i.e., a base station connected to the core network by cable or fiber) using base station-to-relay and relay-to-relay wireless links. With four such relays deployed 200 meters apart, it is able to support cell radius of 900 meters. Three topologies for the wireless backhaul are illustrated here: (a) \emph{Single-Hop} star-topology; (b) \emph{Nearest Neighbor} topology; (c) \emph{full-connectivity} topology.}
		\label{fig:topology}
	\end{figure}

As illustrated in Figure~\ref{fig:topology} we assign one user to each relay/base-station and assume that the distance between each user and its associated relay (or the base station) equals the radius of the associated small cell. Should multiple users arrive at the same relay, the total bandwidth and power can be split to accommodate their service request\footnote{It is to some extend a conservative evaluation of the aggregate cell throughput by putting all users at the edge of their associated small cells (cell radius of 100 meters or less). As we will see later in the simulation results, it is the backhaul rather than the access link that is the bottleneck in achieving high aggregate cell throughput.}.  
 Both backhaul and the access channels use plentiful mmWave spectrum and the relays provide the power boost needed to overcome high path loss between the base station and the users. We assume all relays perform decode-and-forward strategy and have the necessary functionality to handle user access and backhaul communication. 
We further assume that all the backhaul links are orthogonal in bandwidth\footnote{Orthogonal in time is less favorable due to RF peak power constraint.}.  

Wireless backhaul can be established in many different ways. For example, as illustrated in Figure~\ref{fig:topology} (a), the wireless backhaul contains only the direct wireless links between the base station and the relays, which forms a single-hop ``star network'' centered at the base station. Alternatively, wireless backhaul can also be constructed using the ``nearest neighbor'' topology as shown in Figure~\ref{fig:topology} (b). We can also enumerate all possible combinations of direct and relaying links to establish the backhaul, and the case when all the possible wireless links are used for backhaul is illustrated in Figure~\ref{fig:topology} (c).

\subsection{Channel Path Loss Models}

We use four different path loss models which reflect the physical condition of various deployment scenarios. 
The first path loss model is the \emph{blocked Line-of-Sight (LOS)} model proposed in~\cite{CVV2015}, 
where free space path loss is combined with an additional 25 dB shadowing loss to account for possible obstructions.  This path loss model, denoted as \emph{LOS+25dB}, represents a favorable propagation scenario for mmWave transmission and its path loss can be expressed in dB scale as 
	\begin{align}\label{eqn:LOS-25}
		PL_{\mathrm{LOS+25}} = & 20\log_{10}(d)+20\log_{10}(f_c) \nonumber\\
		 & +20\log_{10}(4\pi/0.3)+25, \hspace{1em}\text{[dB]}		
	\end{align}  
	where $ d $ is the distance (in meters) between the transmitter and the receiver and $ f_c $ is the carrier frequency (in GHz).   
	
	Another important model, the 5GCM Urban Macro NLOS (\emph{UMa-NLOS})  ABG model~\cite{5GCM}, represents the deployment scenario of macro base station in urban environment where the base station is placed on high altitude (e.g., a tower of 25 meters) but the direct path from the base station to relays/users are blocked by buildings. Its path loss expression is as follows 
	\begin{align}\label{eqn:UMa-NLOS}
		PL_{\mathrm{UMa-NLOS}}=34\log_{10}(d)+23\log_{10}(f_c)+19.2, \hspace{1em}\text{[dB].}		
	\end{align}
	
The third path loss model we adopt is the 3GPP Urban Micro NLOS
 (\emph{UMi-NLOS})\cite{3GPP} which represents the urban NLOS deployment scenario where the base station is located on a lower altitude (e.g., below clutter) in contrast to the \emph{UMa-NLOS}.  The \emph{UMi-NLOS} path loss model is 
	\begin{align}\label{eqn:UMi-NLOS}
	PL_{\mathrm{UMi-NLOS}}=36.7\log_{10}(d)+26\log_{10}(f_c)+22.7, \hspace{1em}\text{[dB]}.		
	\end{align}
Note that for $f_c=28$ GHz and $d=100$ meters, the path loss of \emph{UMi-NLOS} is 13 dB worse than the path loss of \emph{UMa-NLOS}.
	
To quantify the effect of angular spread on the effective beamforming gain, we use the	3GPP UMi Street Canyon NLOS (\emph{UMi-Street})~\cite{3GPP-above6G} channel model, whose path loss can be written as (assuming user terminal height of 1.5 meters) 
	\begin{align}\label{eqn:UMi-Street}
	PL_{\mathrm{UMi-Street}} = 35.3\log_{10}(d_{3D}) {+} 21.3\log_{10}(f_c) {+} 22.4, \hspace{1mm}\text{[dB]}.
 \end{align}
where $d_{3D}$ is the 3D distance between the transmitter and the receiver. For transmitter height of 10 meters and receiver height of 1.5 meters, separated 100 meters apart, the average RMS azimuth spread  of departure  angle (ASD) as specified in~\cite{3GPP-above6G} is  $15.6^\circ$  and the average RMS zenith spread of departure  angle (ZSD) is  $1.6^\circ$. These angular spread parameters will be used to determine the effective beamforming gain under the	3GPP UMi Street Canyon NLOS model.

 \subsection{Achievable Rate and Channel Estimation Penalty} 

	When the signal to noise ratio (SNR) is high and the small scale fading is mild, the channel estimation penalty is negligible. The  maximum achievable rate  of the channel can be closely approximated by the capacity of continuous time AWGN channel with the same SNR:
	\begin{align}
	R(P_T, W) &= W\log_2(1+\text{SNR})\nonumber\\
		&=W\log_2\left(1+\frac{P_T\cdot PL^{-1}}{N_0\cdot W}\right),\text{ [bits/s]} \label{eqn:shannon}
	\end{align}
  where $ P_T $ denotes the transmitted signal power from the transmitter, $ PL$ is the path loss of the channel in linear scale between the transmitter and the receiver, $ W $ is the allocated bandwidth, and $ N_0 $ is the power spectral efficiency of noise.   
	
	When SNR is low or when the channel changes fast both in time and infrequency, the channel estimation penalty must be taken into account.  Let $L_c$ be the channel coherence length that presents the average number of orthogonal symbols per each independent channel coefficient. For I.I.D. block fading channels with  coherence time $T_c$ and coherence bandwidth $B_c$, the channel coherence length equals the  product of coherence time and coherence frequency, i.e., $L_c=B_cT_c$. By assigning a fraction $\alpha\in(0,1)$ of symbols as pilots and performing minimum mean square error (MMSE) channel estimation at the receiver, as described in~\cite{Hassibi-Hochwald}, the effective SNR will be degraded due to channel estimation error and the amount of resource left for data transmission is reduced to $1-\alpha$. Hence the maximum achievable rate \eqref{eqn:shannon} should be written as~\cite{Hassibi-Hochwald} 
		\begin{align}
	R(P_T, W,\alpha) &= W\log_2(1+ \frac{\alpha L_c \text{SNR}^2}{1+\text{SNR}+\alpha L_c \text{SNR}}), \text{ [bits/s]}, \label{eqn:rate2}
	\end{align}
	where $\alpha L_c$ is the number of pilot symbols that are allocated for channel estimation and $\text{SNR}=\frac{P_T\cdot PL^{-1}}{N_0\cdot W}$ is given in \eqref{eqn:shannon}. 
	By optimizing over\footnote{For any finite channel coherence length $L_c$, we allocate $K$ out of $L_c$ symbols as pilots and therefore the pilot ratio $\alpha=K/L_c$ is a rational number that falls into the range between zero and one. We need to relax $\alpha$ to a real number over the range $(0,1)$ to ease the optimization. This relaxation is mild as long as $L_c$ is large, which is the case in typical wireless communication channels~\cite{BWopt}.}  $\alpha\in(0,1)$ one can find the maximum achievable rate  
	\begin{align}
R(P_T, W) \triangleq \max_{0<\alpha<1} R(P_T, W, \alpha). \label{eqn:Rwa}
\end{align}

\section{Optimal Backhaul Topology and Bandwidth-Power Allocation}\label{sec:optimization}
 
Given the fact that backhaul links can take advantage of all the base station-to-relay and relay-to-relay links, it is natural to ask what is the optimal backhaul topology and associated bandwidth-power partitioning to maximize the user rate or minimum resource usage.
	Instead of iterating all possible heuristic backhaul topology, such as the \textit{Single-Hop} star-topology illustrated in Figure~\ref{fig:topology} (a) and the \textit{Nearest-Neighbor} topology in Figure~\ref{fig:topology} (b), we focus on the full-connectivity topology as shown in Figure~\ref{fig:topology} (c) and  allocate bandwidth and power across all the backhaul links, with the option to not allocate any resource at all to a  specific link.
If the optimization only allocates bandwidth and power to a subset of links, that subset defines the actual backhaul topology. Therefore such optimization, if feasible, will solve the topology-bandwidth-power optimization at the same time. 

However, it is challenging to jointly optimize the topology design and power-bandwidth allocation for the maximum throughput. To establish the relationship between rate and bandwidth, we focus on the dual problem where we minimize the total bandwidth to meet a predefined target data rate for every user in the system. For each target rate, in a steady state, data delivery over the directed graph illustrated in Figure~\ref{fig:topology} (c) can be regarded as a flow network with a single source and multiple sinks. Let $R_n$, $n=1,\ldots,5$, be the target rate for user $n$. Let $t$ be a node in the network (e.g., the base station, a relay, or a user), with data rate $R^t_{in,i}$ on its $i_{th}$ link that delivers data into node $t$ and data rate $R^t_{out,j}$ on its $j_{th}$ link that sends data out. For the flow network in steady state, we have
\begin{align} \label{eqn:flow}
\sum_i R^t_{in,i} - \sum_j R^t_{out,j} = \left\{
\begin{array}{ll}
  - \sum_n R_n, &  t=\text{base station};\\
	0,  & t=\text{relay node};\\
	R_n, & t=\text{user}\ n.
	\end{array}\right.
\end{align}
 We can then decouple the topology design and rate of each channel using linear constraints, and the global optimization problem is now decoupled into sub-problems where the dependence of rate and power-bandwidth for each channel is handled locally in parallel.  

Given some fixed target rates to be satisfied for  users in the system, we try to find the bandwidth allocation vector $\overbar{\bW}$ and power allocation vector $\overbar{\bP}$, where each entry of the vectors corresponds to a wireless backhaul/access link in the system. Let  vector $\overbar{\bR}$  represent the data rate of all the wireless links with bandwidth allocation specified by $\overbar{\bW}$ and  power allocation specified by $\overbar{\bP}$, we have
\begin{align} \label{eqn:rate}
   \overbar{\bR} = f(\overbar{\bW}, \overbar{\bP}),
\end{align} 
where the function $f(\overbar{\bW}, \overbar{\bP})$ denotes the non-linear rate function prescribed by \eqref{eqn:shannon} or \eqref{eqn:Rwa} that is applied element-wise to each channel in the system to obtain the achievable rates for the given transmitted power $\overbar{\bP}$ and bandwidth $\overbar{\bW}$.    

The linear constraints that represent the requirement of data transmission over a flow network, as specified in \eqref{eqn:flow}, will make sure that the input data rate equals the output data rate at every relay node (i.e., non-source non-sink intermediate nodes in a flow network). We can represent them in a matrix format by
\begin{align} \label{eqn:flow-matrix}
  \bA\overbar{\bR}= \bar{\bb},
\end{align}
where matrix $\bA$ and vector $\bar{\bb}$ are determined using the flow network constraint \eqref{eqn:flow} for given target data rates. We specify the  matrix $\bA$ and vector $\bar{\bb}$ for the case with equal target user rates  in Appendix~\ref{sec:Appendix}. 
 The last constraint is  the total power constraint  at any given transmitter (base station or relay). 

The overall optimization problem can now be expressed as
  \begin{equation}
  \begin{aligned}
  & \underset{\overbar{\bW}, \overbar{\bP}}{\text{minimize}}
  & & \overbar{\boldsymbol{1}}^{T}\overbar{\bW},\\
  & \text{subject to}
	&& \bA\overbar{\bR}= \bar{\bb},\\
  &&&  \overbar{\bR} = f(\overbar{\bW}, \overbar{\bP}),\\
  &&&\overbar{\bW}, \overbar{\bP}, \overbar{\bR} \succeq \overbar{\boldsymbol{0}},\\
  &&& \bD \overbar{\bP} \leq \overbar{\boldsymbol{1}}, 
  \end{aligned} \label{eqn:opt}
  \end{equation}
where $\overbar{\boldsymbol{0}}$ and $\overbar{\boldsymbol{1}}$ denote the vectors of all zeros and ones, respectively, of appropriate size, $\bD$ is the matrix for linear power constraint that depends on  power budget at each transmitter. An example of $\bD$ with separate power constraints for backhaul and for access is given in Appendix~\ref{sec:Appendix}.

The constrained optimization problem defined in \eqref{eqn:opt} has a linear objective function, three non-negativity constraints on optimizing variables, a linear equality constraint  originated from the flow network as specified in \eqref{eqn:flow}, a non-linear equality constraint \eqref{eqn:rate} on the rate of each wireless channel, and a linear inequality on the transmit power. Although the analytical solution for the optimization problem defined in \eqref{eqn:opt} is out of reach, it can be solved efficiently using numerical optimization thanks to the nice property of the problem. As discussed earlier in this section, the non-linear equality rate constraint \eqref{eqn:rate} is enforced element-wise: it represents many parallel rate constraints that are either defined by \eqref{eqn:shannon} or \eqref{eqn:Rwa}. Furthermore, for given power $P$ (resp. bandwidth $W$) the rate \eqref{eqn:shannon} is a concave function of bandwidth $W$ (resp. power $P$)~\cite{Verdu02}. For given $P$ and $W$, \eqref{eqn:rate2} is a concave function of the pilot ratio $\alpha$ and the pilot optimization in \eqref{eqn:Rwa} can be solved very efficiently~\cite{Hassibi-Hochwald}. Therefore the non-linear equality constraint on the rate of each individual channel can be handled locally in parallel, and the overall optimization problem defined in \eqref{eqn:opt} can be solved efficiently, for example, using penalty and barrier methods~\cite{Wright2006} by transferring it to unconstrained optimization using a barrier-penalty function derived from the equality and inequality constraints. In fact, the optimization problem defined in \eqref{eqn:opt}  can also be solved efficiently 
 using standard numerical optimization toolboxes (for example, the MATLAB Optimization toolbox).

We find numerical solutions for different path loss models, and compare them against the results obtained using the \textit{single-hop} and the \textit{nearest-neighbor} topologies. Note that the joint topology-bandwidth-power optimization naturally yields the best result, and the optimized power and bandwidth allocation may turn off some backhaul links in the \textit{full-connectivity} topology to minimize the resource usage. This means, under some deployment scenarios, simpler backhaul topologies such as the \textit{single-hop} and the \textit{nearest-neighbor} schemes may approach/achieve the optimum rate.

\section{Effective Beamforming Gain Limited by Channel Angular Spread}\label{sec:angular}

In ideal case where we have as many RF chains as the antenna elements and we have perfect channel state information, generalized beamforming will provide full array gain, in absolute value,  that grows linearly with the number of antenna elements, unless the size of the array is smaller than the beam itself. However, in practice the number of RF chains are limited due to hardware and cost constraints, and perfect channel state information is not available. Therefore the full array gain indicated by generalized beamforming is out of reach. Instead, beam-steering approach is used to harvest the beamforming gain.  

Given limited number of RF chains and non-zero channel angular spread, as the number of antenna elements increases, the effective beamforming gian will saturate at the limit imposed by the angular spread of the channel. We consider the case of high gain antennas, assumed to have Gaussian shaped beams both in azimuth and elevation to keep the treatment tractable. For an antenna of RMS elevation beamwidth $B_v$ and RMS azimuth beamwidth $B_h$, the antenna gain pattern assumes the form
  \begin{align} \label{eqn:beam}
		g(\phi,\theta)= \frac{2}{B_h B_v}  e^{- \frac{\phi^2}{2 B_h^2 }}   e^{- \frac{\theta^2}{2B_v^2}}. 
	\end{align}

In the absence of scattering, for an array of $N$ elements, each with (linear in power) directional gain (i.e., the element gain)  $G_e$, the maximum antenna gain $g_{\text{max}}$  is related to the antenna RMS beamwidths through
\begin{align} \label{eqn:gain}
   g_{\text{max}}   = NG_e = \frac{2}{B_{h} B_{v}}, 
	\end{align}
where the RMS elevation beamwidth $B_v$ and RMS azimuth beamwidth $B_h$ are set, respectively, to the nominal beamwidth $B_{v0}$ and  $B_{h0}$, respectively, as measured in an anechoic chamber.  
This allows determination of the effective beamwidths of such arrays based on maximum gain. 

In the presence of channel angular spread, induced by scattering, the effective antenna pattern is given by a convolution of the ideal antenna pattern and the channel power angular spectrum. Assuming, for tractability, a Gaussian channel angular spectrum of azimuthal angular spread $\sigma_h$ and elevation angular spread  $\sigma_v$, the resulting effective antenna pattern still has the Gaussian form as above but with effective beamwidths  given by
\begin{align} \label{eqn:beamwidth} 
 B_v =\sqrt{B_{v0}^2+\sigma_v^2 }, \    B_h=\sqrt{B_{h0}^2+\sigma_h^2 }.
\end{align}
 These equations hold approximately for narrow antenna beams and angle spreads of interest. When such spreads begin to approach $\pi/2$, more generalized equations may be derived using the same approach, to assure proper normalization of the effective antenna pattern above.

\begin{figure}[t]
		\centering
		\includegraphics[width=0.99\columnwidth]{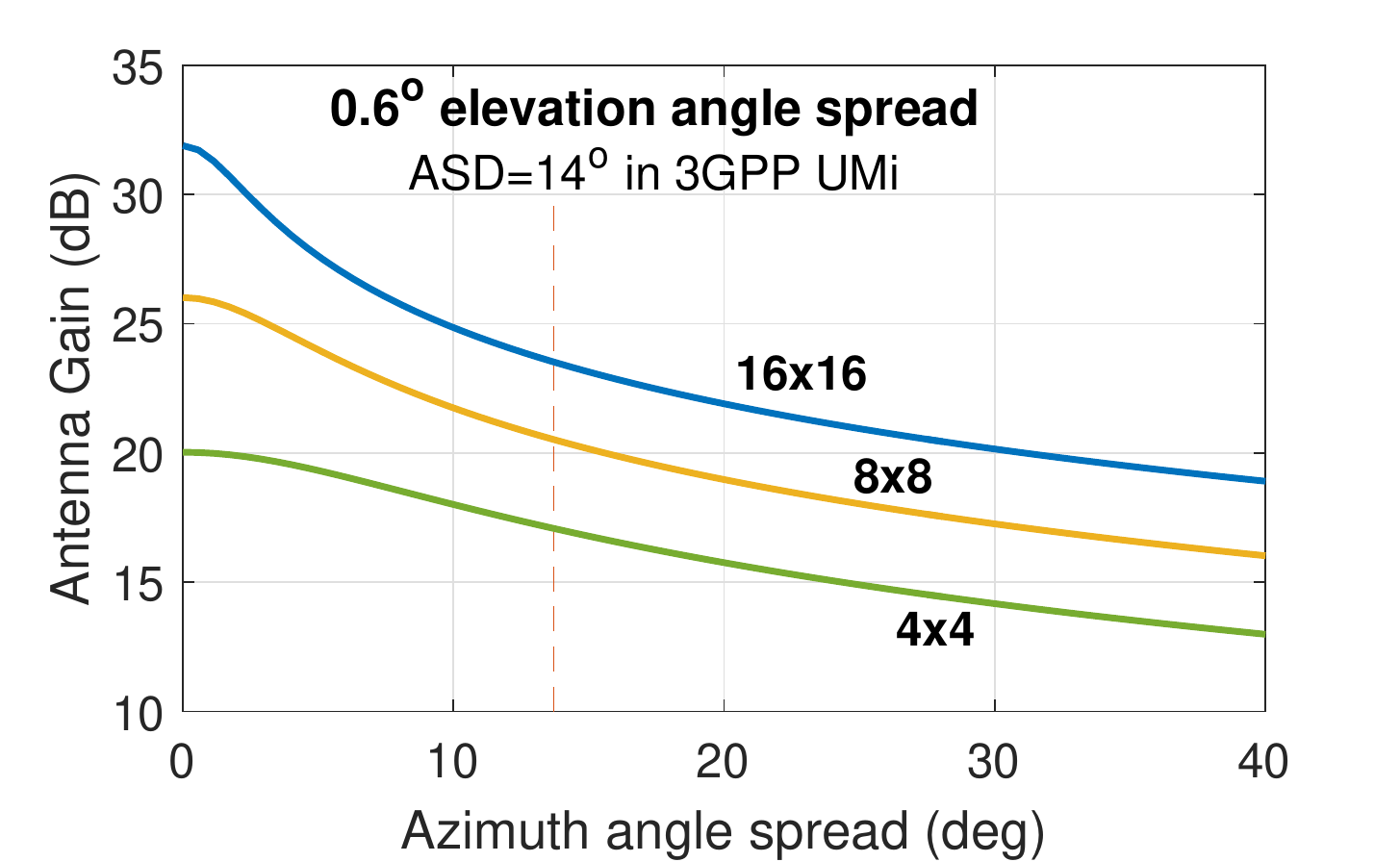}
		\caption{Effective antenna gain as a function of the RMS azimuth angle spread (ASD) with a fixed elevation angle spread  of 0.6 degree. The arrays sizes are as indicated, with each element having 8 dBi gain. The dashed orange vertical line indicates the ASD of 14 degree for the 3GPP Urban Micro Street Canyon LOS path loss model, where we see roughly 9 dB  gain degradation for large array of $16\times16$ and  3 dB gain degradation for small array of size $4\times4$.}
		\label{fig:angular}
	\end{figure}

     To quantify the effective antenna gain degradation with angle spread, in Figure~\ref{fig:angular} we plot the effective antenna gain as a function of the RMS  azimuth angle spread (ASD) for the 28GHz band under 3GPP Urban Micro Street Canyon LOS path loss model~\cite{3GPP-above6G} using a fixed RMS elevation angle spread of 0.6 degree. The arrays sizes are as indicated, with each element having 8 dBi gain.  With ASD of 14 degree as specified in~\cite{3GPP-above6G}, we see roughly 9 dB antenna gain degradation for large array of $16\times16$ and 3 dB gain degradation for small array of size $4\times4$.		
		Should the elevation angle spread  prove to be higher, say, 10 degree instead of 0.6 degree as specified in~\cite{3GPP-above6G}, the antenna gain will saturate much faster.

\section{Numerical Results}\label{sec:num}

\subsection{System Setup}

Since the mmWave links over a few hundred meters are usually limited by power, the benefit of transmitting multiple data streams (i.e., spatial multiplexing) in a single link is very limited. Therefore in the simulation we adopt the  beam-switching transmission proposed in~\cite{CVV2014} where the transmitter and the receiver first sweeping over all the candidate beamforming vectors out of a pre-determined codebook, and then the receiver, after measuring all the beamformed channels, finds the ``hottest'' beam and sends its index to the transmitter for data transmission. The delay and cost of feedback is not considered here.  

To highlight the  influence of path loss, antenna gains, angular spread, array size, and RF power limitation,
we focus on the linear network illustrated in Figure~\ref{fig:topology}, where four relays are deployed along the street, 200 meters apart,  on one side of the base station. All the relays can connect to the base station and other relays using wireless backhaul links, and therefore the backhaul links have distance of 200, 400, 600, and 800 meters. We assign one user to each assess node (e.g., base station or relay) and put the user's position at 100 meters away from their corresponding access points to represent cell edge connections. 
Should the relays positioned at irregular distances or not along a line, or the users are located at various distances, the corresponding link distance should be used in the simulation. 
 
 	 All backhaul links are kept orthogonal in frequency (i.e., no reuse), with distant interference from other base stations or out-of-cell relays neglected.
	For access links,  interference from neighboring cells is removed through frequency reuse of 2, with more distant interference neglected. Should power leakage to adjacent frequency bands become an issue, allocation of  frequency can be arranged in such a way that adjacent frequency bands are allocated to links that are physically separated. The potential of frequency reuse among backhaul links and the associated opportunity for cooperation among relays are left to future study.

For the simulation in this work, we consider the downlink scenario and set identical target rates across all users to simplify the results presentation. We do not resort to polarization, which has the potential to halve the required bandwidth. The simulation parameters are chosen, unless stated otherwise,  as specified in Table~\ref{tab:para}. The carrier frequency is set to $f_c = 28$ GHz which is one of the frequency bands reserved for mmWave communication~\cite{fcc}.

\begin{table}[t]
	\centering
		\begin{tabular}{|c|c|}
		\hline
Transmit power at    &	1 Watt for Access Link\\
  base station/relays   &    1 Watt for Backhaul links\\ \hline
Joint transmit-receive   &	25 dBi for Access link\\ 
 antenna gain &  50 dBi for Backhaul links \\ \hline
Access range	& 100 meters (users at cell edge)   \\ \hline
Backhaul ranges	& 200/400/600/800 meters \\ \hline
Polarization	& Single polarization \\ \hline
Frequency for Access	& Orthogonal, with reuse 2  \\ \hline
 Frequency for Backhaul &  Orthogonal, no reuse\\ \hline
Path loss  & as in \eqref{eqn:LOS-25}, \eqref{eqn:UMa-NLOS}, \eqref{eqn:UMi-NLOS}, \eqref{eqn:UMi-Street}\\ \hline 
Carrier frequency & $f_c=28$ GHz \\ \hline
Noise figure & 9 dB \\ \hline
		\end{tabular}
		\vspace{1mm}
	\caption{Common Simulation Parameters}
	\label{tab:para}
\end{table}

\subsection{Impact of Joint Antenna Gains}

	\begin{figure}[t]
		\centering
		\includegraphics[width=0.98\columnwidth]{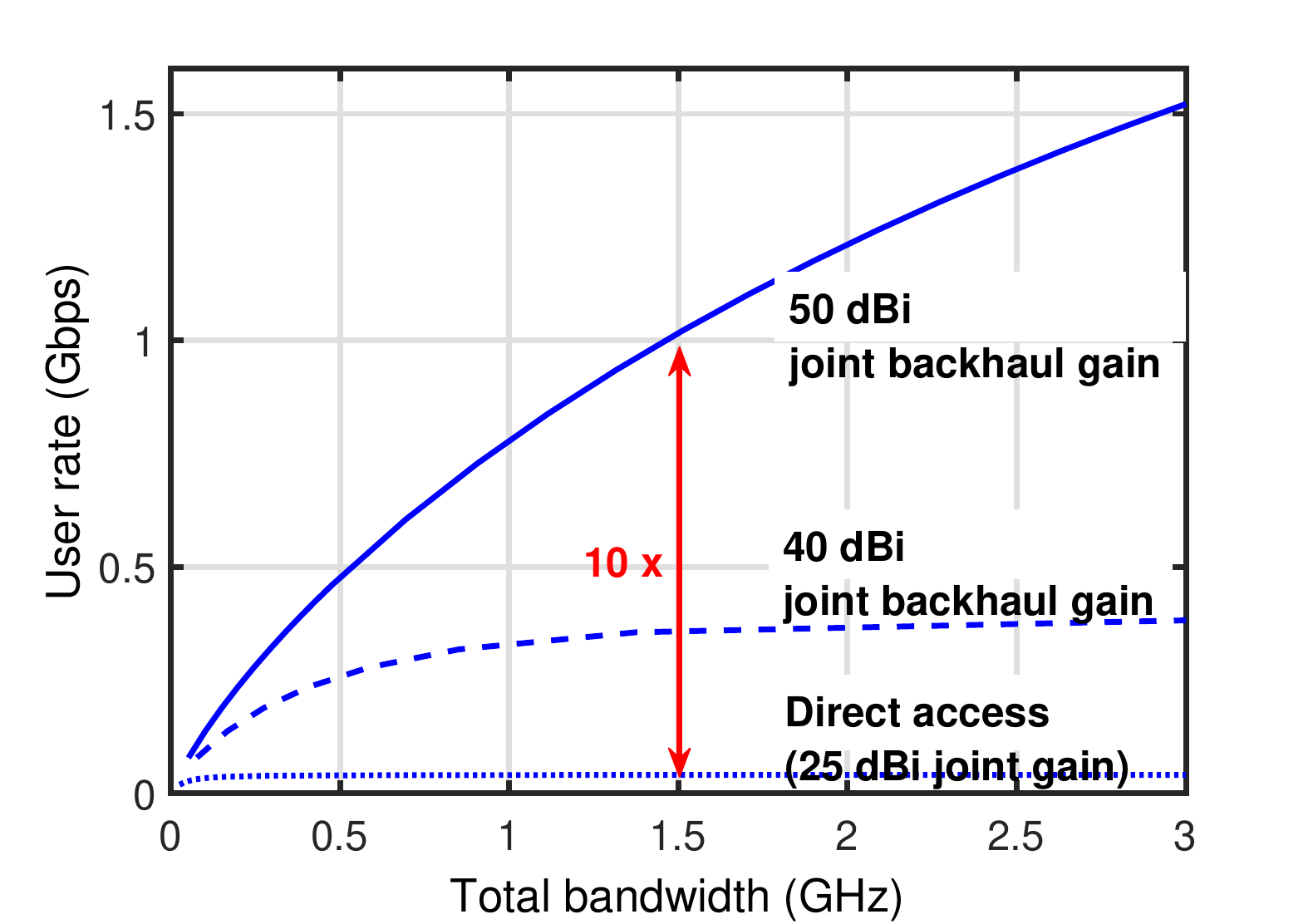}
		\caption{Per user throughput (in total 5 users) as a function of total bandwidth (Backhaul $+$ Access) for the \textit{single-hop} ``star-network'' backhaul topology. With high directional gain of 50 dBi for backhaul links provides 1Gbps per user throughput, which is more than 10 times faster than direct access  (i.e., no relaying) where the joint transmit-receive antenna gain is assumed to be 25 dBi. If the combined array gain for backhaul links falls short to 40dBi due to scattering, the rate gain falls to less than 400 Mbps/user.}
		\label{fig:gain}
	\end{figure}	
	
In Figure~\ref{fig:gain} we plot the per user throughput (in total 5 users) as a function of total bandwidth for the \textit{single-hop} ``star-network'' backhaul topology, where the power is equally split among the four backhaul links (base station to relay), i.e., 250 mW for each link. With ideal joint  transmit-receive antenna gain of 50 dBi, the throughput reaches 1 Gbps/user with 1.5 GHz total bandwidth consumption. This is more than 10 times rate gain compared to the direct access scenario (i.e., no relaying) where joint transmit-receive antenna gain is 25 dBi. However, if the materialized joint antenna gain falls short to 40 dBi, for example, due to scattering, the rate gain falls to less than 400 Mbps/user.  Such rate loss cannot be remedied through using higher bandwidth. This demonstrates the critical importance of using high gain antennas in supporting Gbps user rate over traditional cell range.

\subsection{Impact of Channel Path Loss}

	\begin{figure}[t]
		\centering
		\includegraphics[width=0.98\columnwidth]{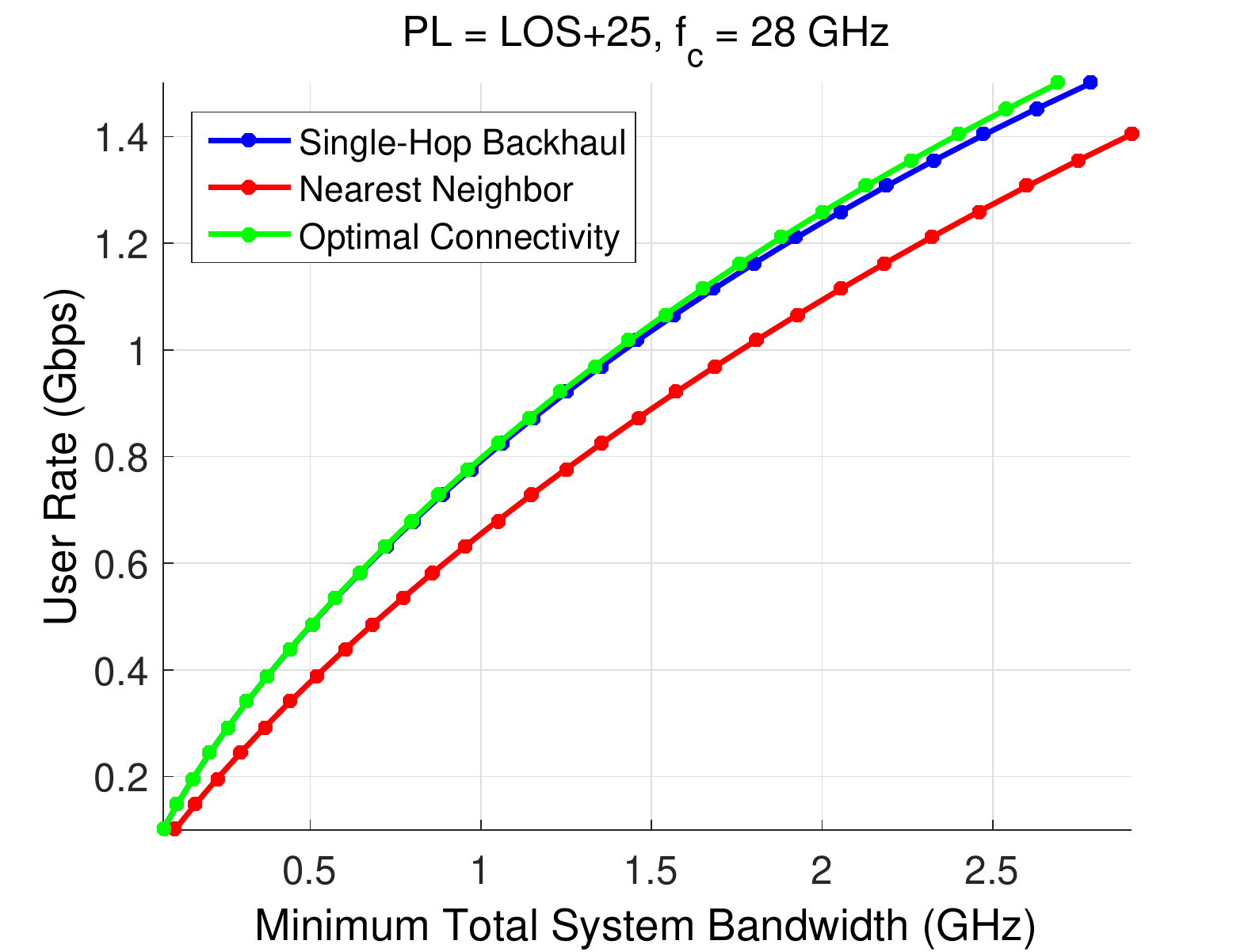}
		\caption{Per user throughput as a function of the total bandwidth for the \emph{LOS+25dB} path loss model. The curve labeled as ``Optimal Connectivity'' refers to the results we have obtained from joint topology-bandwidth-power optimization.   Since the path loss is favorable, the \textit{Single-Hop} backhaul is within 2\% from the optimum whereas the \textit{Nearest-Neighbor} suffers 15\%-20\% rate loss.}
		\label{fig:LOS25}
	\end{figure}	

\begin{figure}[t]
		\centering
		\includegraphics[width=0.98\columnwidth]{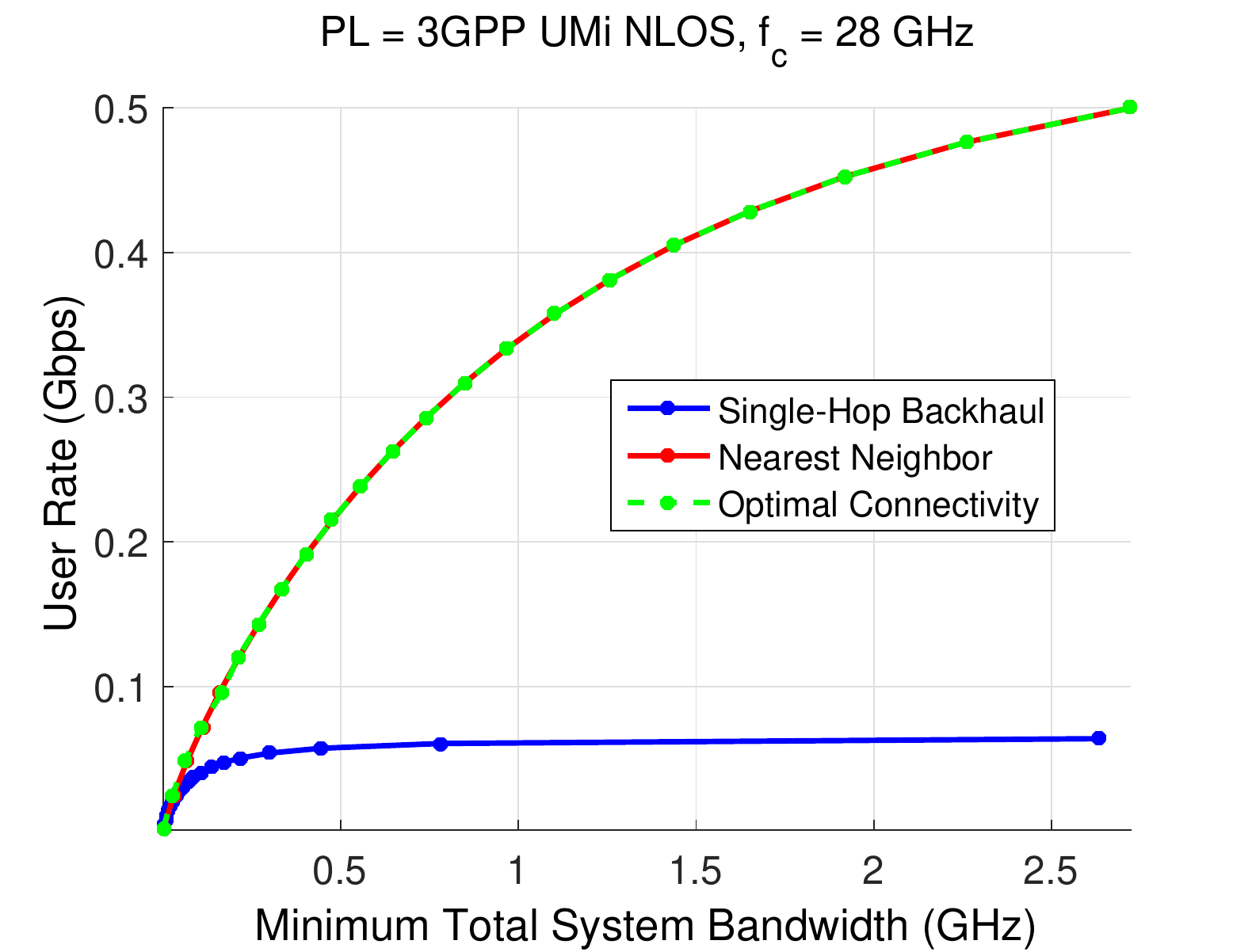}
		\caption{Per user throughput as a function of the total bandwidth for the \emph{UMi-NLOS} path loss model. Since the path loss is severe, long distance transmission is not favorable and the \textit{Nearest-Neighbor} backhaul performs almost as good (or not-so-good) as the jointly optimized backhaul. 		
		The \textit{Single-Hop} backhaul, on the other hand, fails to deliver rate higher than 70 Mbps/user.}
		\label{fig:UMi}
	\end{figure}

In Figure~\ref{fig:LOS25} we plot the per user throughput as a function of the total bandwidth for the \emph{LOS+25dB} path loss model, where the curve labeled as ``Optimal Connectivity'' refers to the results we have obtained from joint topology-bandwidth-power optimization. The curves obtained using optimized bandwidth and power for the \textit{Single-Hop} backhaul topology and the \textit{Nearest-Neighbor} topology are plotted as reference.  When the path loss is favorable, which is the case in the \emph{LOS+25dB} path loss model, the \textit{Single-Hop} backhaul is within 2\% from the optimum whereas the \textit{Nearest-Neighbor} suffers 15-20\% rate loss.
When the path loss is severe, as in the case for the \emph{UMi-NLOS} path loss model shown in Figure~\ref{fig:UMi}, long distance transmission is severely attenuated. The \textit{Nearest-Neighbor} backhaul performs   as good  as the jointly optimized backhaul.	
		The \textit{Single-Hop} backhaul, on the other hand, fails to deliver rates higher than 70 Mbps/user.

The results presented in Figure~\ref{fig:LOS25} and Figure~\ref{fig:UMi} reveal an interesting interplay between the target user data rate, path loss models, and the optimal backhaul topology. Firstly, Optimal Connectivity always gives the best result among all schemes, as expected. Secondly, the \textit{Single-Hop} backhaul topology is very close to optimum  under favorable path loss as in the \emph{LOS+25dB} path loss model  while it becomes severely non-optimal in \textit{UMi-NLOS} model. On the other hand, the \textit{Nearest-Neighbor} backhaul topology has a considerable gap (~15\%--20\%) from the optimal scheme in the \textit{LOS+25 } models while it is the optimal scheme when long distance transmission is highly discriminated by the channel as in the \textit{UMi-NLOS} model. This suggests that as the path loss becomes more severe, there is a topological transition from the \textit{Single-Hop} to \textit{Nearest-Neighbor} to achieve the optimum.

\subsection{Impact of Joint Topology-Power-Bandwidth Optimization}

\begin{figure}[t]
		\centering
		\includegraphics[width=0.98\columnwidth]{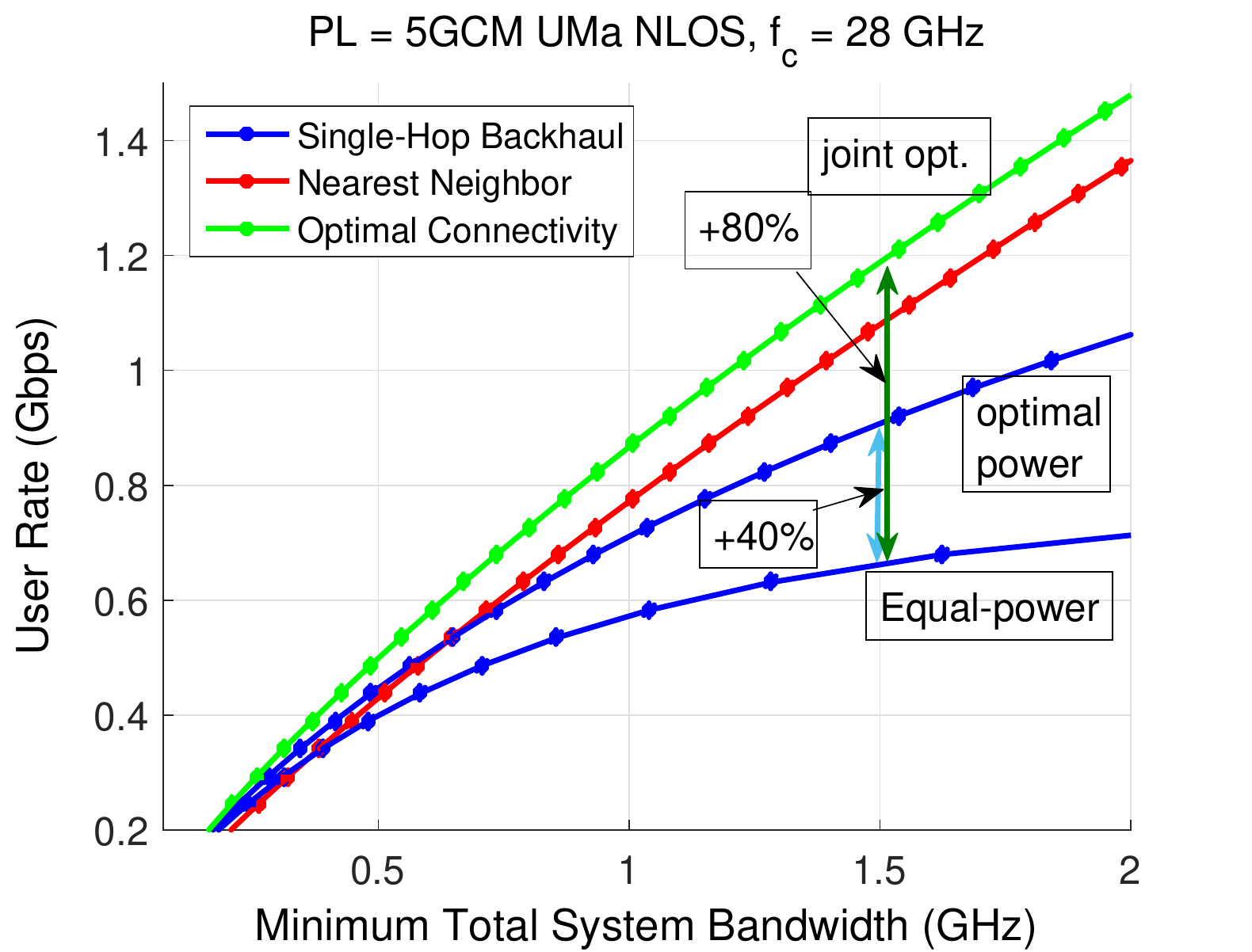}
		\caption{Per user throughput as a function of the total bandwidth for the \emph{UMa-NLOS} path loss model. The curve using the \textit{Single-Hop} backhaul but with equal power split is also plotted as reference to indicate the benefit (larger than 40\% rate gain in this case) of power optimization. Joint topology-power-bandwidth optimization will increase the rate gain to more than 80\%.}
		\label{fig:UMa-gain}
	\end{figure}

To highlight the importance of joint topology-power-bandwidth optimization, in
Figure~\ref{fig:UMa-gain} we compare the per user rates achieved under the \textit{UMa-NLOS} model by using the
\textit{Single-Hop} backhaul with equal power split, \textit{Single-Hop} with optimized power allocation, and the joint topology-power-bandwidth optimization.  With the \textit{ Single-Hop} backhaul topology, power optimization alone brings more than 40\% rate gain. The rate gain grows to more than 80\%  if we also optimize the topology. 

\begin{figure}[t]
		\centering
		\includegraphics[width=0.98\columnwidth]{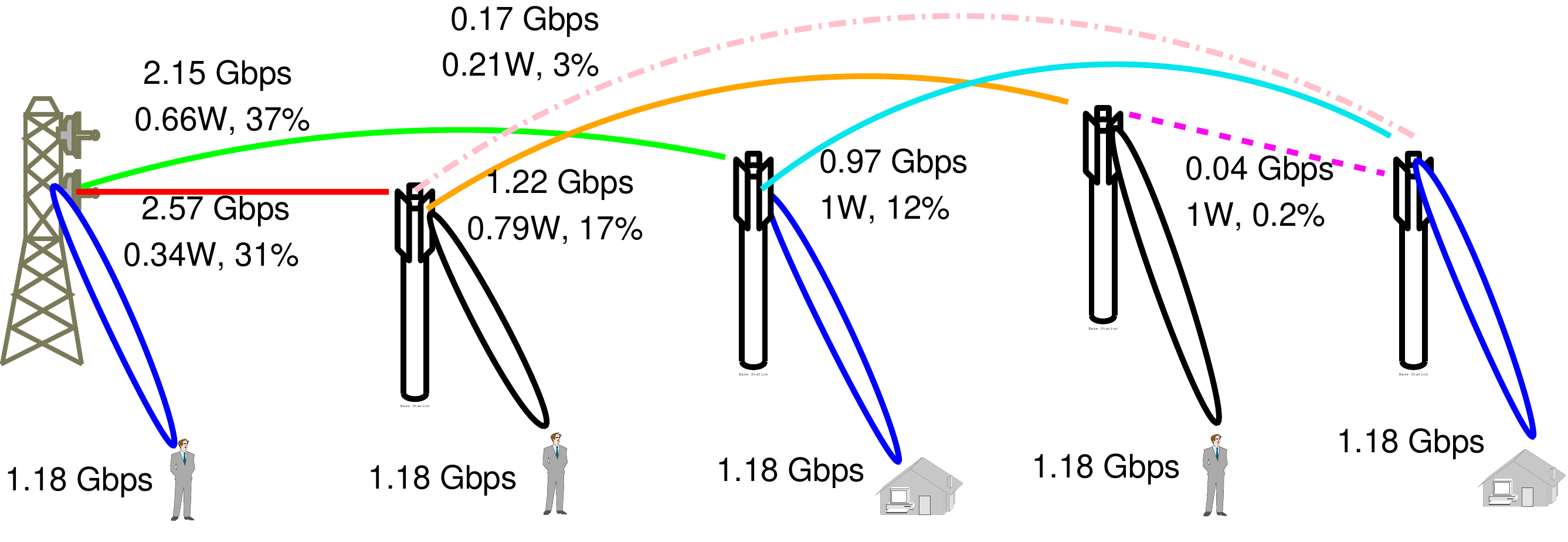}
		\caption{The optimized backhaul topology and associated power-bandwidth allocation to provide 1 Gbps/user rate for the \emph{UMa-NLOS} path loss model as shown in Figure~\ref{fig:UMa-gain}. We label each of the backhaul links with the corresponding rate, power (in Watt), and bandwidth (in percentage).  The percentage indicates the share of backhaul link bandwidth out of the total backhaul bandwidth 1049 MHz whereas two frequency bands of 219 MHz each are allocated to Access. Therefore, 1487 MHz of total system bandwidth are used to support the 1.18 Gbps/user rate. Note that all numbers are rounded at the last digit for labeling. }
		\label{fig:Opt-topology}
	\end{figure}

A close inspection on how the optimized backhaul topology and associated power-bandwidth allocation work together will provide us valuable insights for the design of wireless backhaul in supporting Gbps/user rate. In Figure~\ref{fig:Opt-topology} we label each backhaul link by the its rate\footnote{For convinence of labeling, all numbers for rate/power/bandwidth are rounded to the last digit as shown in the figure.}, allocated power (in Watt), and bandwidth (in percentage) for the case where 1 Gbps/user rate is supported under the  \emph{UMa-NLOS} path loss model. The percentage corresponds to the ratio of allocated bandwidth to the total bandwidth of the backhaul system (in this case, 1049 MHz). Together with the two Access bands, each of 219 MHz, the total system bandwidth is 1487 MHz to support the 1.18 Gbps/user rate.   Note that the last two relays are connected to the base station only through the help of the first two relays. This topology is not close to either of the two heuristic backhaul topologies we have compared with and it necessitates the joint optimization prescribed in \eqref{eqn:opt}.

\subsection{Impact of Array Size and Antenna Element Power}

\begin{figure}[t]
		\centering
		\includegraphics[width=0.98\columnwidth]{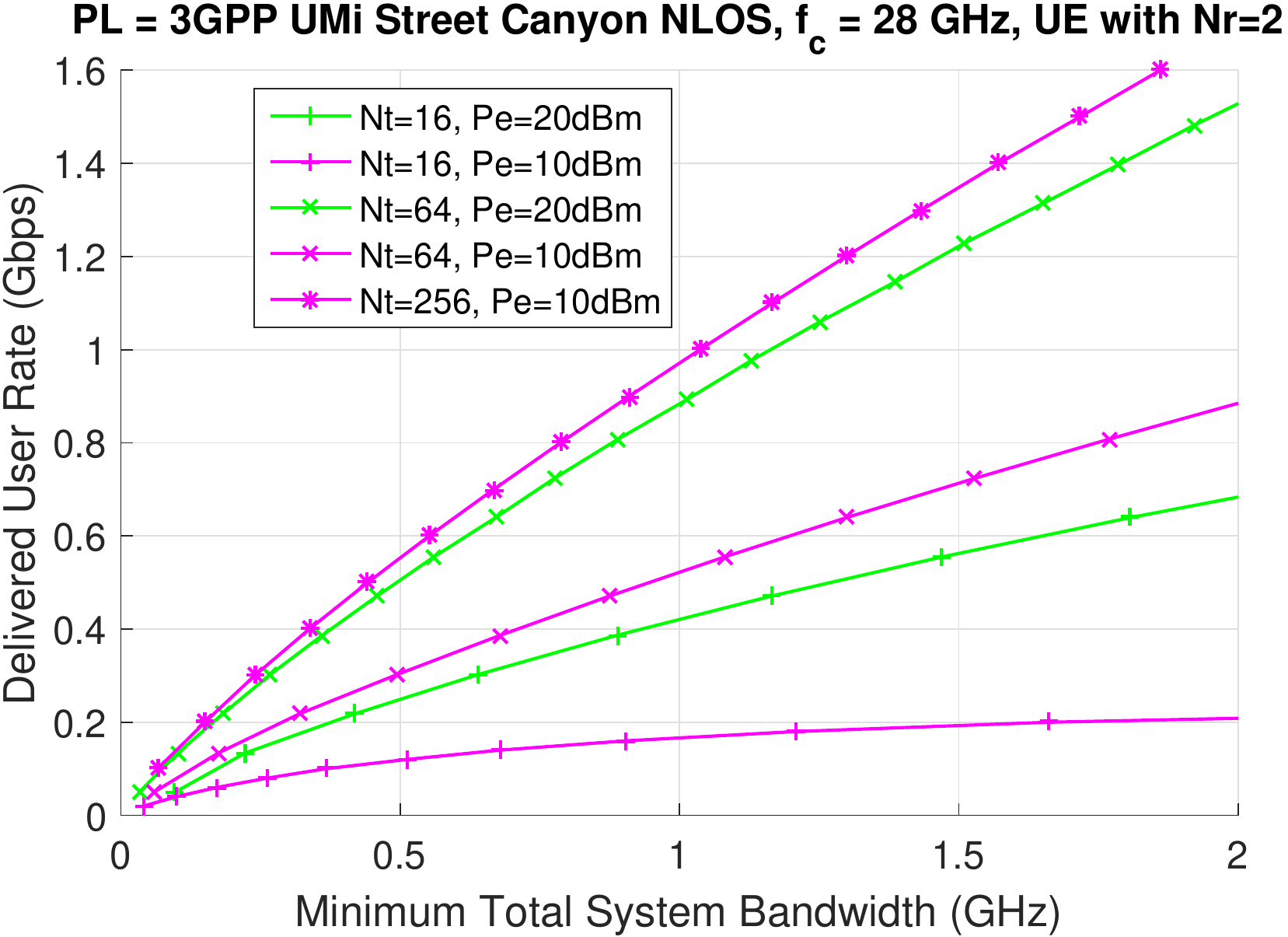}
		\caption{ Per user throughput as a function of the total bandwidth for the  3GPP \textit{UMi Street Canyon NLOS} path loss model~\cite{3GPP-above6G}  with coherence time $T_c=5$ ms and coherence bandwidth $B_c=7.6$ MHz. Transmit array size changes from (16, 64, 256) and per-element power is chosen from (10 dBm, 20 dBm), each with 8 dBi element gain. Each user is equipped with $N_r=2$ antenna elements, each of 5 dBi element gain. When increasing the array size at the users to $N_r=8$, the curves will almost overlap with those for $N_r=2$ and therefore omitted.}
		\label{fig:UMiStr-PeNt}
	\end{figure}

In all the previous plots, we have assumed 1 Watt transmit power and 25 dBi antenna gain at the base station and the relays, without considering their dependence. Since channel estimation cost grows with array size and the effective beamforming gain saturates due to angular spread, the  impact of antenna array configuration on the overall performance should be investigated. In Figure~\ref{fig:UMiStr-PeNt} we plot the rate-bandwidth curves as a function of the transmit array size (16, 64, 256) and the per-element power (10 dBm, 20 dBm). The channel path loss and angular spread are taken from the 3GPP \textit{UMi Street Canyon NLOS} model~\cite{3GPP-above6G} with coherence time $T_c=5$ ms (assuming roughly 100 Hz Doppler shift) and coherence bandwidth $B_c=7.6$ MHz. Each user is equipped with $N_r=2$ antenna elements, each of 5 dBi element gain. The results show that, with 1.2 GHz total bandwidth, if 20 dBm per-element power is available, 64-element antenna array at the base station and the relays  would be sufficient to provide 1 Gbps/user rate for users at cell edge. If the per-element power is only 10 dBm\footnote{With 10 dBm per-element power, an array of 100 elements will provide RF power of 1 Watt.}, array size of 256 elements is needed to support similar user rates. However, increasing the array size at the user side does not help much to improve the downlink user rate since it is the Backhaul rather than the Access that is the bottleneck for high speed downlink mmwave access over long distance.

\section{Conclusions}\label{sec:conclusion}

In this work,  we have proposed a generic framework to design wireless relayed backhaul by jointly optimizing its backhaul topology,  power, and bandwidth. This is done by focusing on the dual problem and by using the flow network  to decouple the backhaul topology and bandwidth-power constraints for each relay node. We have evaluated the influence of path loss, angular spread,  array size, and RF power limitation on the user rate through numerical simulations under different path loss models. It has been shown that for a linear network deployed along the street at 28 GHz, 1 Gbps/user  rate within cell range of 1 km can be delivered using 1.5 GHz  bandwidth and high-gain  single-polarization antennas. The desired total bandwidth can be potentially halved by adopting dual-polarization antennas. The user rates drop precipitously when joint directional gain is reduced, or when the path loss is much more severe. When the number of RF chains is limited, the effective beamforming gain saturates owing to the channel angular spread. When the per-element power is high (20 dBm), Gbps/user rate can be delivered by deploying 64-element antenna arrays at the base station and the relays. If the per-element gain is only 10 dBm, the desired array size is of 256 elements. The benefit of larger arrays will eventually be surpassed by the increased channel estimation penalty due to beamforming gain saturation. Although the numerical results presented here are customized for linear networks, which is most suitable for street canyon scenarios, our analysis framework and the optimization setup can be extended straightforwardly to more general relay networks by adjusting the link distances in the optimization setup.

There are many ways to extend the current work. Frequency reuse among backhaul links has the potential to reduce the required bandwidth to support Gbps user rates at cell edge. Since one user's data may be available in its two neighboring relays, a byproduct of inter-relay backhaul links, it opens room for network coding or cooperative transmission schemes to further increase user rate. On the other hand, the inter-cell interference, which may arise from different relays within the same macro cell due to frequency reuse, or from other relays in the neighboring macro cell when two beams pointing to each other along the boresight. When the power of such interference is not very small, the signal to interference plus noise ratio should be used rather than the SNR. The corresponding bandwidth and pilot optimization may incur some dependence among different links. Should the framework be extended to traditional cell structure, the position of the relays and the potential of frequency reuse among backhaul links will also play a major role. We leave this to future research.   

\section*{Acknowledgment}

  We would like to thank Dr. Greg Wright and Dr. Shahriar Shahramian for discussions on latest advances in RF power engineering.
 
\appendices
 
\section{Optimization Formulation}\label{sec:Appendix}

\begin{figure}[t]
		\centering 
		\includegraphics[width=0.96\columnwidth]{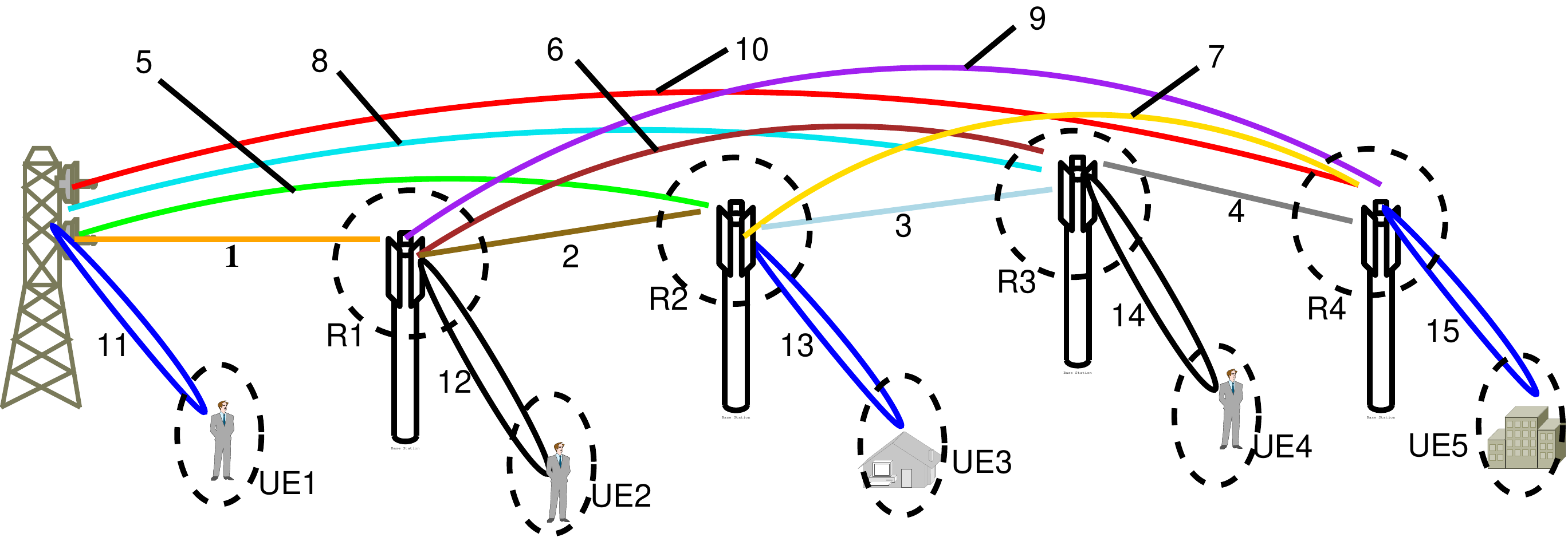} 
		\caption{Illustration of the flow network for the \emph{full-connectivity} topology stated in Figure~\ref{fig:topology} (c) where the backhaul links are numbered from 1 to 10 and the five access links are numbered from 11 to 15. Each dashed circle corresponding to a constraint imposed by the flow network as specified in  \eqref{eqn:flow}. Wireless link $n$, $n=1,\ldots, 15$, is represented by the $n$th column of matrix $\bA$ and the flow constraint on relay $k$, $k=1,\ldots, 4$, is represented by the $k$th row of matrix $\bA$. The flow constraint on UE $m$, $m=1,\ldots, 5$, is represented by the $(m+4)$th row of matrix $\bA$.}
		\label{fig:flow}
	\end{figure}

We denote $ \overbar{\bW} $ and $ \overbar{\bP} $  the vectors of bandwidths and powers allocated to each channel. Each vector has as many entries as the number of channels in the system. As an example, for the full-connectivity scenario illustrated in Fig.~\ref{fig:topology},
we number the wireless links as illustrated in Figure~\ref{fig:flow}. The index of the elements in each vector $ \overbar{\bW} $ and $ \overbar{\bP} $ refers to the channels labeled by the same number: $ W_1 $ (resp., $ P_1 $) is the bandwidth (resp., power) allocated to the channel between the base station and the first (closest to base station) relay, $ W_2 $ (resp., $ P_2 $) is the bandwidth (resp., power) for the channel between the first and second relay, and so on. Five access channels constitute the last elements of $ \overbar{\bW} $ and $ \overbar{\bP} $.  The objective function of the optimization problem given in \eqref{eqn:opt} is the sum of bandwidths allocated to all channels in the system. Naturally all the bandwidths and powers must be non-negative which is reflected by the second constraint. The  constraint  $ \overbar{\bR} = f(\overbar{\bW}, \overbar{\bP}) $ on the vector of data rates $ \overbar{\bR} $ achieved in each channel  is a function of bandwidth and power allocated to that channel  as given by \eqref{eqn:shannon} or \eqref{eqn:Rwa}. The constraint $\bA\overbar{\bR}=\bar{\bb}$ ensures that, in the absence of relay cooperation, the sum of the data rates that go into a node must be the same as that goes out, unless it is the base station or a user, as specified in \eqref{eqn:flow}. Thus for the full-connectivity backhaul topology with a target per user rate of $ R^* $, we can set $ \bar{\bb}=\overbar{\boldsymbol{1}} $ and rewrite the flow constraint  as $ \bA  \overbar{\bR} =\overbar{\boldsymbol{1}} $, where the $n$th column of matrix $\bA$ represents the wireless link $n$, $n=1,\ldots, 15$, and  the rows of $\bA$ represent the flow constraint \eqref{eqn:flow}: the first four rows refer to the flow constraints on relays 1 to 4, and the remaining five rows refer to the flow constraints on user 1 to 5. Therefore we can write the matrix $\bA$ as shown in \eqref{eqn:mA}, at the top of the next page.  
\textcolor{black}{Note that the flow constraint on the base station is not explicitly shown in matrix $\bA$, but has been implicitly imposed: when adding up the first four rows of $\bA$  we will obtain a row vector with ``$1$'' on the 1st, 5th, 8th, and 10th columns and ``$0$'' otherwise, and on the right hand side of the equation the sum of the first four elements of $\overbar{\boldsymbol{1}}$ equals 4.}

 \begin{figure*}
\begin{equation}\label{eqn:mA}
	\bA=\frac{1}{R^*}\left[
	\begin{array}{ccccccccccccccc}
	 1& -1& 0& 0& 0& -1& 0& 0& -1& 0& 0& 0& 0& 0& 0\\
	 0& 1& -1& 0& 1& 0& -1& 0& 0& 0& 0& 0& 0& 0& 0\\
	 0& 0& 1& -1& 0& 1& 0& 1& 0& 0& 0& 0& 0& 0& 0\\
	 0& 0& 0& 1& 0& 0& 1& 0& 1& 1& 0& 0& 0& 0& 0\\
	 0& 0& 0& 0& 0& 0& 0& 0& 0& 0& 1& 0& 0& 0& 0\\
	 0& 0& 0& 0& 0& 0& 0& 0& 0& 0& 0& 1& 0& 0& 0\\
	 0& 0& 0& 0& 0& 0& 0& 0& 0& 0& 0& 0& 1& 0& 0\\
	 0& 0& 0& 0& 0& 0& 0& 0& 0& 0& 0& 0& 0& 1& 0\\
	 0& 0& 0& 0& 0& 0& 0& 0& 0& 0& 0& 0& 0& 0& 1\\
	\end{array}\right].
\end{equation}  
 \end{figure*}

Finally, the power constraint on each transmitter is reflected by the inequality $  \bD \overbar{\bP}=\overbar{\boldsymbol{1}} $  in the optimization formulation. We have assumed that there are separate power budget on backhaul and access channels. For instance, with power budget $ P_b $ for Backhaul  and $ P_a $ for Access at each relay node and at the base station, $\bD $ takes the following form for the full-connectivity network,  shown as \eqref{eqn:MD} at the top of the next page.

  \begin{figure*}
\begin{equation}\label{eqn:MD}
\bD =\left[\begin{array}{ccccccccccccccc}
\frac{1}{P_b}& 0& 0& 0& \frac{1}{P_b}& 0& 0& \frac{1}{P_b}& 0& \frac{1}{P_b}& 0& 0& 0& 0& 0\\
0& \frac{1}{P_b}& 0& 0& 0& \frac{1}{P_b}& 0& 0& \frac{1}{P_b}& 0& 0& 0& 0& 0& 0\\
0& 0& \frac{1}{P_b}& 0& 0& 0& \frac{1}{P_b}& 0& 0& 0& 0& 0& 0& 0& 0\\
0& 0& 0& \frac{1}{P_b}& 0& 0& 0& 0& 0& 0& 0& 0& 0& 0& 0\\
0& 0& 0& 0& 0& 0& 0& 0& 0& 0& \frac{1}{P_a}& 0& 0& 0& 0\\
0& 0& 0& 0& 0& 0& 0& 0& 0& 0& 0& \frac{1}{P_a}& 0& 0& 0\\ 
0& 0& 0& 0& 0& 0& 0& 0& 0& 0& 0& 0& \frac{1}{P_a}& 0& 0\\ 
0& 0& 0& 0& 0& 0& 0& 0& 0& 0& 0& 0& 0& \frac{1}{P_a}& 0\\ 
0& 0& 0& 0& 0& 0& 0& 0& 0& 0& 0& 0& 0& 0& \frac{1}{P_a}
\end{array}\right].
\end{equation}
 \end{figure*}

\begin{IEEEbiography}[{\includegraphics[width=1.05in,height=1.1in, clip,keepaspectratio]{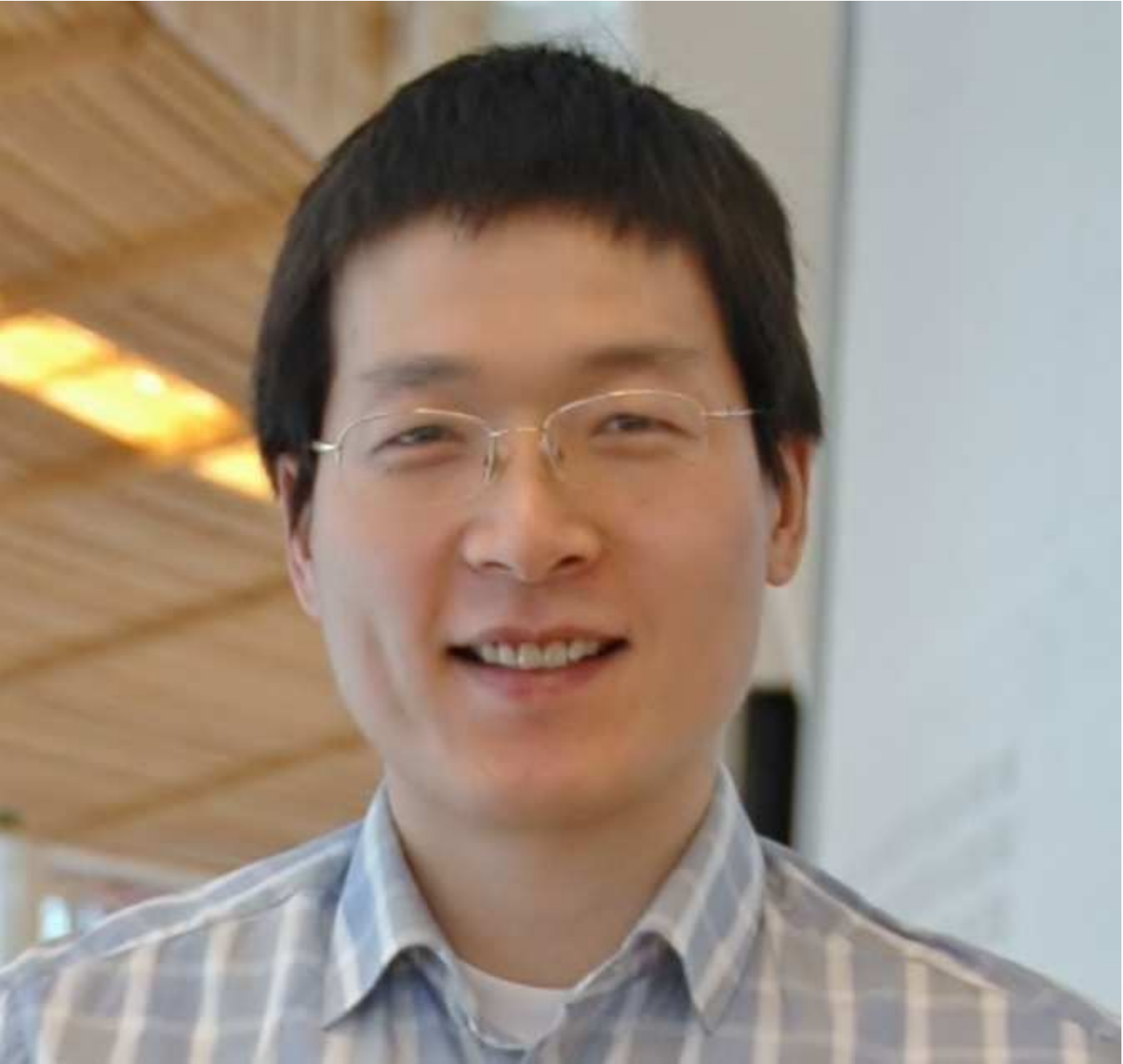}}]{Jinfeng Du}
(S’07-M'13) received his B.Eng. degree in electronic information engineering from the University of
Science and Technology of China (USTC), Hefei, China, and the M.Sc., Tekn. Lic., and Ph.D. degrees from the Royal Institute of Technology (KTH), Stockholm, Sweden. He was a postdoctoral researcher at the Massachusetts Institute of Technology (MIT), Cambridge, MA, from 2013 to 2015, after which he joined Bell Labs at Crawford Hill, Holmdel, NJ, where he is currently a Member of Technical Staff. His research interests are in the general area of wireless communications, communication theory,  information theory,  and wireless networks. Dr.~Du received the Best Paper Award from IC-WCSP in October 2010, and his paper was elected as one of the ``Best 50 Papers'' in IEEE GLOBECOM 2014. He received the prestigious ``Hans Werth\'en Grant''  from the
Royal Swedish Academy of Engineering Science (IVA) in 2011, the ``Chinese Government Award for
Outstanding Self-Financed Students Abroad'' in 2012, and  the ``International PostDoc'' grant from
the Swedish Research Council in 2013. He also received three grants from the Ericsson Research Foundation.
\end{IEEEbiography}

\begin{IEEEbiography}[{\includegraphics[width=1in,height=1.1in, clip,keepaspectratio]{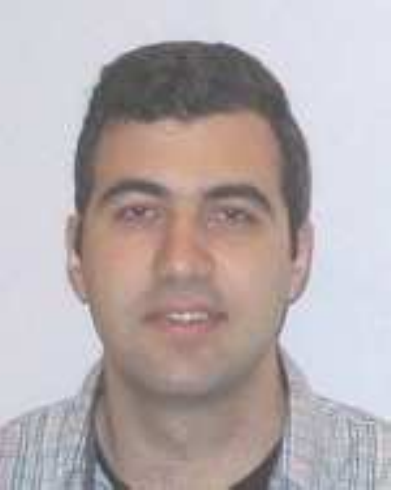}}]{Efe Onaran} received the B.S. degree in electrical engineering from Bilkent University, Ankara, Turkey, in 2013 and M.S.E. degree from Princeton University, NJ, USA, in 2015. He is currently working toward the Ph.D. degree in electrical engineering at New York University Tandon School of Engineering, Brooklyn,
NY, USA. His research interests include theory, algorithms and applications of random graphs and optimization.
\end{IEEEbiography}

\begin{IEEEbiography}[{\includegraphics[width=1in,height=1.1in, clip,keepaspectratio]{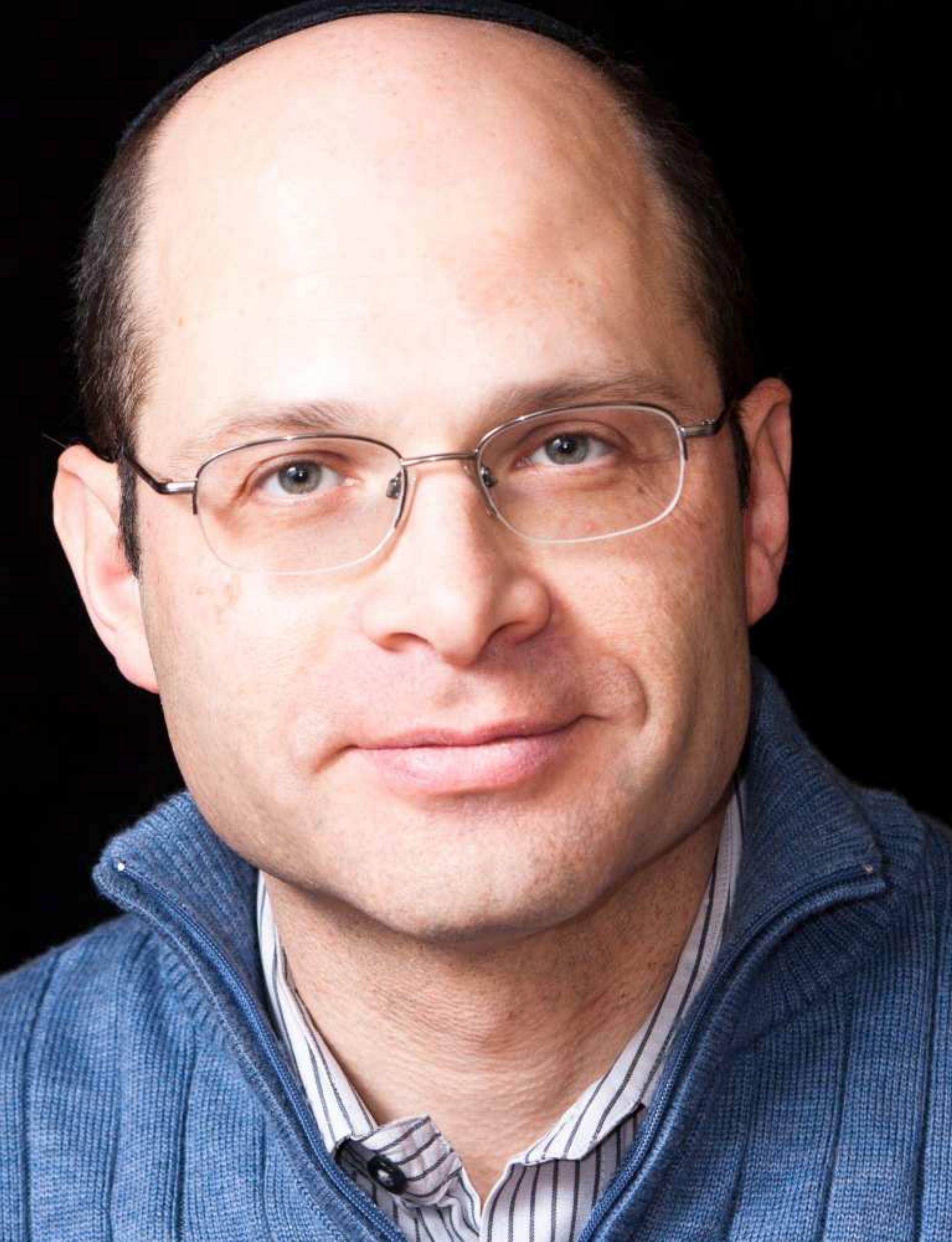}}]{Dmitry Chizhik} (F'14) received the Ph.D. degree in electrophysics from Polytechnic University (now NYU), Brooklyn, NY, USA. His thesis work has been in ultrasonics and non-destructive evaluation. He joined the Naval Undersea Warfare Center, New London, CT, USA, where he did research in scattering from ocean floor, geoacoustic modeling of porous media and shallow water acoustic propagation. In 1996, he joined Bell Labs, where he is involved in radio propagation modeling and measurements, using deterministic and statistical techniques.  The results are used both for determination of channel-imposed bounds on channel capacity, system performance and for optimal antenna array design. His recent work has included system and link simulations of satellite and femto cell radio communications and mm wave propagation covering all aspects of the physical layer. His research interests are in acoustic and electromagnetic wave propagation, signal processing, communications, radar, sonar, and medical imaging. He a recipient of the Bell Labs President's Award.
\end{IEEEbiography}
 
\vspace{13cm}
\newpage 

\begin{IEEEbiography}[{\includegraphics[width=1in,height=1.1in, clip,keepaspectratio]{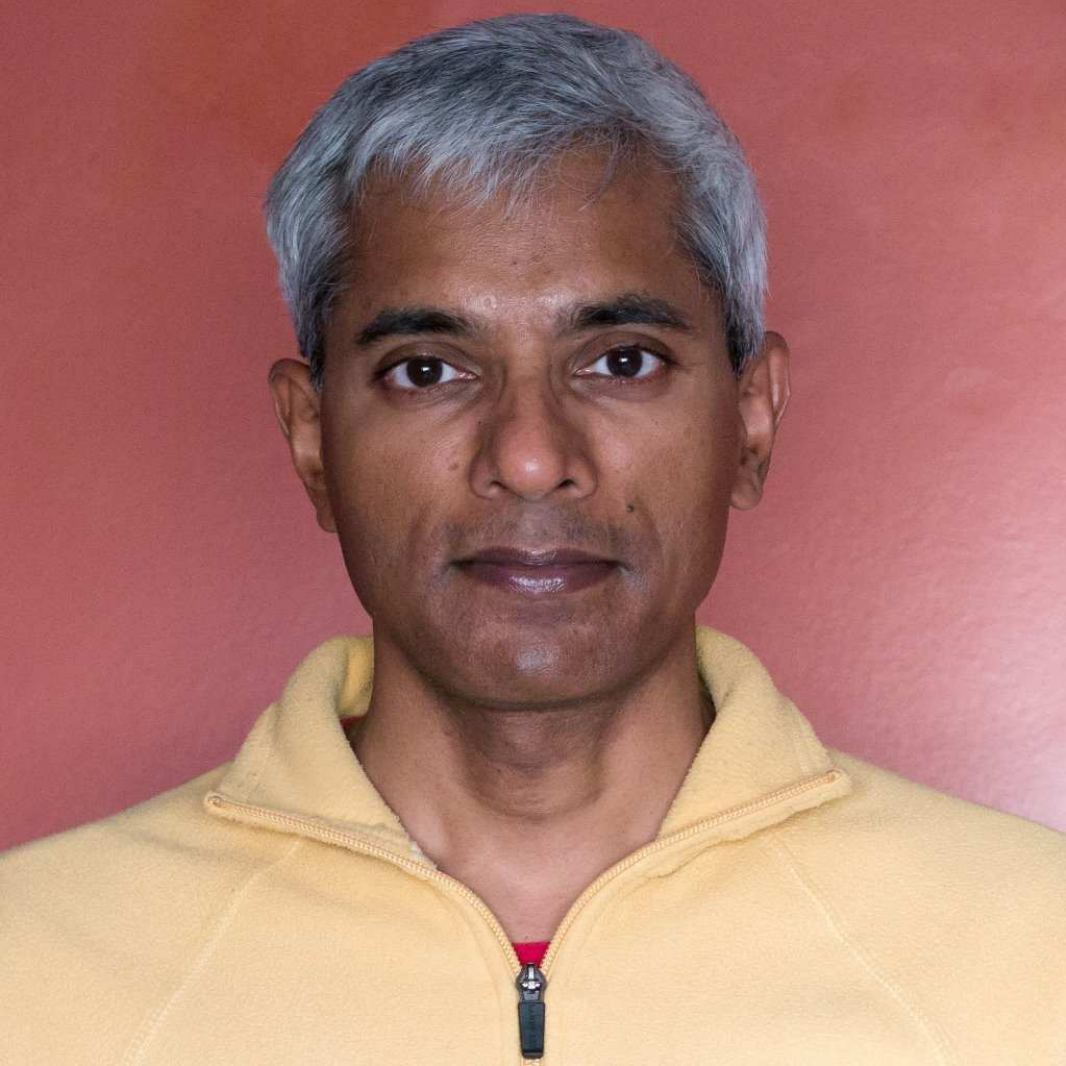}}]{Sivarama Venkatesan} received his Ph.D. in Electrical Engineering from Cornell University, Ithaca, NY, U.S.A., in 1998, and has been with Bell Labs, NJ, U.S.A., since 2001. His research here has been on MIMO techniques for cellular systems, load balancing and interference coordination in heterogeneous networks, multi-carrier waveforms for 5G wireless, and millimeter-wave system performance evaluation, among other areas.  He is a co-author of ``MIMO Communication for Cellular Networks,'' published by Springer in 2012.
\end{IEEEbiography}

\begin{IEEEbiography}[{\includegraphics[width=1in,height=1.1in, clip,keepaspectratio]{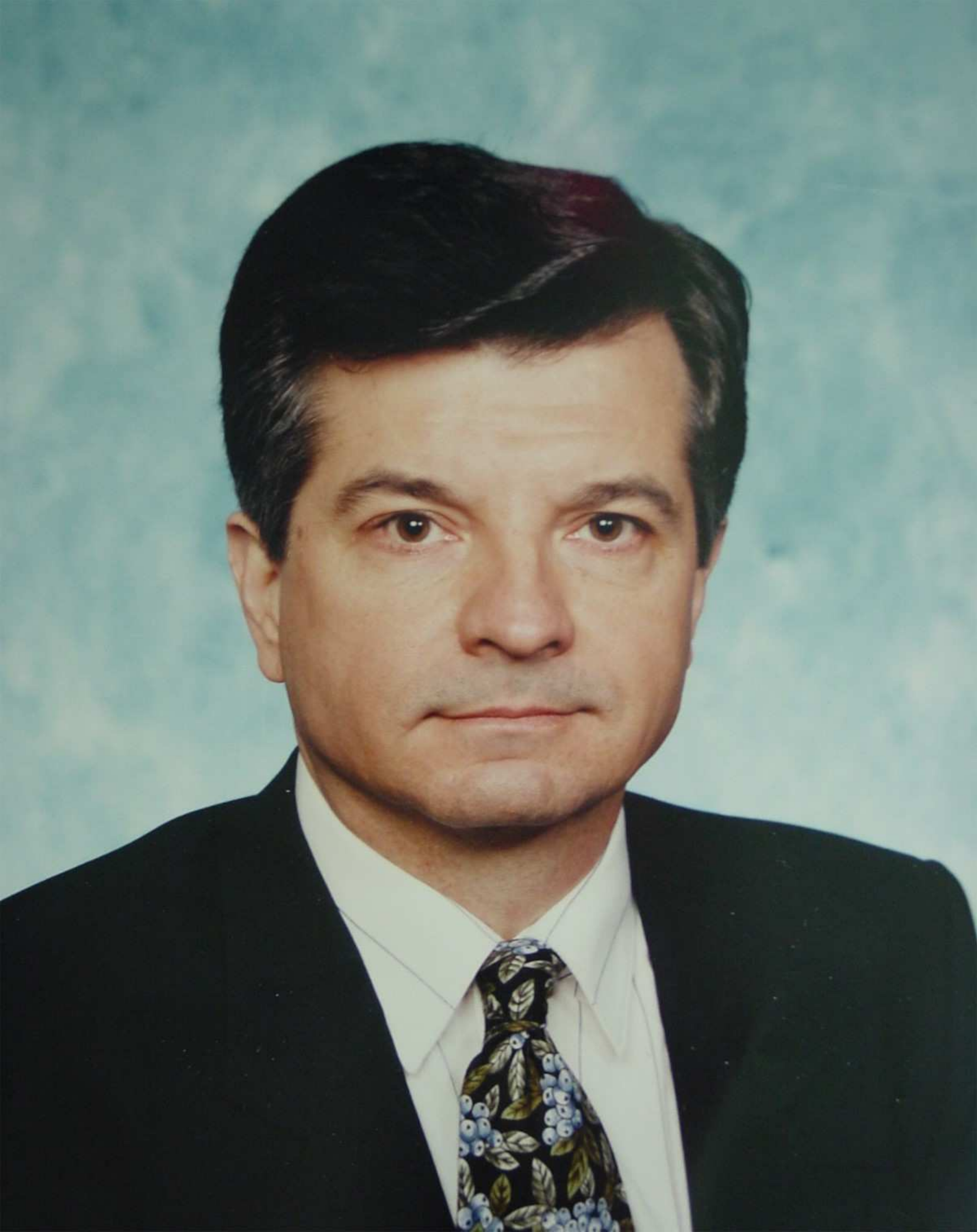}}]{Reinaldo A. Valenzuela} (M'85–SM'89–F'99) received the B.Sc. degree from the University of Chile, Santiago, Chile, and the Ph.D. degree from the Imperial College London, London, U.K. He is currently the Director of the Communication Theory Department and a Distinguished Member of Technical Staff with Bell Laboratories, Crawford Hill, NJ, USA. He is currently engaged in propagation measurements and models, MIMO/space time systems achieving high capacities using transmit and receive antenna arrays, HetNets, small cells and next generation air interface techniques and architectures. 

Dr. Valenzuela is a member of the US National Academy of Engineering and a Fellow of the IEEE. He is a Bell Labs Fellow, a WWRF Fellow, and a Fulbright Senior Specialist. He was a recipient of the 2010 IEEE Eric E. Sumner Award, the 2014 IEEE CTTC Technical Achievement Award, and the 2015 IEEE VTS Avant Garde Award. He has published 190 papers and 44 patents. He has over 26,000 Google Scholar citations and is a 'Highly Cited Author' In Thomson ISI.
\end{IEEEbiography}

\vspace{13cm} \hspace{1mm}


\begin{thebibliography}{99}


\bibitem{fcc} \emph{Report and Order and Further Notice of Proposed Rulemaking}, Federal Communications Commission, FCC-16-89, July 14, 2016. 

\bibitem{Hassibi-Hochwald} B. Hassibi and B. M. Hochwald, ``How much training is needed in multiple-antenna wireless links,'' \emph{IEEE Transactions on Information Theory}, vol.~49, pp.~951--963, Apr. 2003.

\bibitem{BWopt} J. Du and R. A. Valenzuela, ``How much spectrum is too much in millimeter wave wireless access,''  \emph{IEEE Journal on Selected Areas in Communications}, accepted for publication.

		
\bibitem{Pi2011} Z. Pi and F. Khan, ``An introduction to millimeter-wave mobile broadband
systems,'' \emph{IEEE Communications Magazine}, vol.~49, pp.~101--107, Jun. 2011.

\bibitem{Rangan2014} S. Rangan, T. Rappaport, and E. Erkip, ``Millimeter-wave cellular wireless
networks: Potentials and challenges,'' \emph{Proceedings of IEEE}, vol.~102, 
pp.~366--385, Mar. 2014.

 \bibitem{Ghosh2014} A. Ghosh et al., ``Millimeter wave enhanced local area systems: A high
data rate approach for future wireless networks,'' \emph{IEEE Journal on Selected Areas in Communications},  vol.~32, pp.~1152--1163, Jun. 2014. 


	
\bibitem{maric2004bandwidth}
I.~Maric and R.~D. Yates, ``Bandwidth and power allocation for cooperative
  strategies in Gaussian relay networks,'' in \emph{proceedings of the Thirty-Eighth Asilomar Conference
  on Signals, Systems and Computers}, vol.~2, pp.  1907--1911, 2004.

\bibitem{reznik2002capacity}
A.~Reznik, S.~R. Kulkarni, and S.~Verdu, ``Capacity and optimal resource
  allocation in the degraded Gaussian relay channel with multiple relays,'' in
  \emph{proceedings of the Annual Allerton Conference on Communication Control
  and Computing}, vol.~40,  pp. 377--386, 2002.

\bibitem{islam2012optimal}
M.~N. Islam, N.~Mandayam, and S.~Kompella, ``Optimal resource allocation and
  relay selection in bandwidth exchange based cooperative forwarding,'' in
  \emph{proceedings of the 10th International Symposium on Modeling and Optimization in Mobile, Ad Hoc and Wireless Networks  (WiOpt)}, pp. 192--199, 2012.

\bibitem{bakanoglu2011resource}
K.~Bakanoglu, S.~Tomasin, and E.~Erkip, ``Resource allocation for the parallel
  relay channel with multiple relays,'' \emph{IEEE Transactions on Wireless
  Communications}, vol.~10,  pp. 792--802, Mar. 2011.

\bibitem{liang2004resource}
Y.~Liang and V.~V. Veeravalli, ``Resource allocation for wireless relay
  channels,'' in \emph{proceedings of the Thirty-Eighth Asilomar Conference on Signals, Systems and Computers}, vol.~2, pp. 1902--1906, 2004.

\bibitem{liang2007resource}
Y.~Liang, V.~V. Veeravalli, and H.~V. Poor, ``Resource allocation for wireless
  fading relay channels: Max-min solution,'' \emph{IEEE Transactions on
  Information Theory}, vol.~53,  pp. 3432--3453, Oct. 2007.

\bibitem{gong2010joint}
X.~Gong, S.~A. Vorobyov, and C.~Tellambura, ``Joint bandwidth and power
  allocation in wireless multi-user decode-and-forward relay networks,'' in
  \emph{proceedings of the IEEE International Conference on Acoustics, Speech and Signal
  Processing}, pp. 2498--2501, 2010.

\bibitem{zhao2006improving}
Y.~Zhao, R.~Adve, and T.~J. Lim, ``Improving amplify-and-forward relay
  networks: optimal power allocation versus selection,'' in \emph{proceedings of IEEE
  International Symposium on Information Theory}, pp. 1234--1238, 2006.

\bibitem{gao2008channel}
F.~Gao, T.~Cui, and A.~Nallanathan, ``On channel estimation and optimal
  training design for amplify and forward relay networks,'' \emph{IEEE
  Transactions on Wireless Communications}, vol.~7,  pp. 1907--1916,
  May 2008.
 
\bibitem{Hur2013} S. Hur, T. Kim, D. J. Love, J. V. Krogmeier, T. A. Thomas, and A. Ghosh, ``Millimeter wave beamforming for wireless backhaul and access in small cell networks,'' \emph{IEEE Transactions on 
  Communications}, vol. 61, pp. 4391--4403, Oct. 2013.

\bibitem{Taori2015} R. Taori and A. Sridharan, ``Point-to-multipoint in-band mmwave backhaul for 5G networks,'' \emph{IEEE Communications Magazine}, vol.~53,  pp.~195--201, Jan. 2015.		 

\bibitem{Singh2015}	S. Singh, M. N. Kulkarni, A. Ghosh, and J. G. Andrews, ``Tractable
model for rate in self-backhauled millimeter wave cellular networks,''
 \emph{IEEE Journal on Selected Areas in Communications}, vol.~33,  pp.~2196--2211, Jan. 2015.


\bibitem{biswas2016performance}
S.~Biswas, S.~Vuppala, J.~Xue, and T.~Ratnarajah, ``On the performance of relay
  aided millimeter wave networks,'' \emph{IEEE Journal of Selected Topics in
  Signal Processing}, vol.~10,  pp. 576--588, Apr. 2016.


\bibitem{CVV2015} D. Chizhik, R. A. Valenzuela, and S. Venkatesan, ``Physical limits on beam switching performance of LOS mmWave links,'' \emph{Bell Labs Technical Report}, {ITD-15-55823C}, May 2015. 

\bibitem{5GCM} ``5G channel model for bands up to 100 GHz,'' \emph{5GCM White paper}, Dec. 2015. \url{http://www.5gworkshops.com/5GCM.html}

\bibitem{3GPP} ``Further advancements for E-UTRA physical layer aspects,'' \emph{3GPP Technical Report TR 36.814 v9.0.0}, Mar. 2010. \url{http://www.qtc.jp/3GPP/Specs/36814-900.pdf} 


\bibitem{3GPP-above6G} ``Study on channel model for frequency spectrum above 6 GHz,'' \emph{3GPP Technical Report TR 38.900 v14.1.0},   Sep. 2016. 
\url{http://www.3gpp.org/ftp/specs/archive/38_series/38.900/38900-e10.zip}


\bibitem{CVV2014} D. Chizhik, S. Venkatesan, and R. A. Valenzuela, ``Viability of mmWave spectrum for outdoor, hot-spot small cells,'' \emph{Bell Labs Technical Report}, {ITD-14-54831Z}, Mar. 2014.


	\bibitem{Verdu02}
S.~Verd\'{u}, ``{Spectral efficiency in the wideband regime},'' \emph{IEEE
  Transactions on Information Theory}, vol.~48, pp. 1319--1343, Jun. 2002.

\bibitem{Wright2006} J. Nocedal and S.~J.~Wright, \emph{Numerical Optimization, 2nd Edition}, Springer-Verlag, New York, 2016.


\end{thebibliography}
\end{document}